\documentclass[twocolumn]{aastex631} 

\usepackage{verbatim}
\usepackage{longtable}

\begin{document}

\title{Sensitive 3~mm Imaging of Discrete Sources in the Fields of X-ray-Selected Galaxy Clusters}

\correspondingauthor{Simon Dicker}
\email{sdicker@physics.upenn.edu}
\author[0000-0002-1940-4289]{Simon R Dicker}
\affiliation{Department of Physics and Astronomy, University of Pennsylvania, 209 S. 33rd St., Philadelphia, PA 19014, USA}
\author[0000-0002-3169-9761]{Mark Devlin}
\affiliation{Department of Physics and Astronomy, University of Pennsylvania, 209 S. 33rd St., Philadelphia, PA 19014, USA}
\author[0000-0003-3586-4485]{Luca Di Mascolo}
\affiliation{Kapteyn Astronomical Institute, University of Groningen, Landleven 12, 9747 AD, Groningen, The Netherlands}
\author[0000-0001-6519-502X]{Saianeesh Haridas}
\affiliation{Department of Physics and Astronomy, University of Pennsylvania, 209 S. 33rd St., Philadelphia, PA 19014, USA}
\author[0000-0002-8490-8117]{Matt Hilton}
\affiliation{Wits Centre for Astrophysics, School of Physics, University of the Witwatersrand, Private Bag 3, 2050, Johannesburg, South Africa}
\affiliation{School of Mathematics, Statistics, and Computer Science, University of KwaZulu-Natal, Westville Campus, Durban 4041, South Africa}
\author[0000-0001-9830-3103]{Joshiwa van Marrewijk}
\affiliation{Leiden Observatory, Leiden University, P.O. Box 9513, 2300 RA Leiden, The Netherlands}
\author[0000-0002-8472-836X]{Brian S. Mason}
\affiliation{National Radio Astronomy Observatory, 520 Edgemont Rd., Charlottesville, VA 22903, USA}
\author[0000-0001-9793-5416,gname=Emily,sname=Moravec]{Emily Moravec}
\affiliation{Green Bank Observatory, P.O. Box 2, Green Bank, WV 24944}
\author[0000-0003-3816-5372]{Tony Mroczkowski}
\affiliation{Institute of Space Sciences (ICE-CSIC), Carrer de Can Magrans, s/n, 08193 Cerdanyola del Vallès, Barcelona, Spain}
\affiliation{Institut d'Estudis Espacials de Catalunya (IEEC), E-08860 Castelldefels, Barcelona, Spain}
\author[0000-0003-1842-8104]{John Orlowski-Scherer}
\affiliation{Department of Physics and Astronomy, University of Pennsylvania, 209 S. 33rd St., Philadelphia, PA 19014, USA}
\author[0000-0001-5725-0359]{Charles Romero}
\affiliation{Department of Physics and Astronomy, University of Pennsylvania, 209 S. 33rd St., Philadelphia, PA 19014, USA}
\affiliation{ADNET Systems, Inc., 6720B Rockledge Drive, Bethesda, MD 20817, USA}
\author[0000-0003-0167-0981,gname=Craig,sname=Sarazin]{Craig Sarazin}
\affiliation{Department of Astronomy and Virginia Institute for Theoretical Astronomy, University of Virginia, P.O. Box 400325, Charlottesville, VA 22904, USA}
\author[0000-0001-6903-5074]{Jonathan Sievers}
\affiliation{Department of Physics, McGill University, 3600 University Street Montreal, QC, H3A 2T8, Canada}

\begin{abstract}
We present the results of a blind survey for compact sources in 138 galaxy clusters from the eFEDS X-ray survey. Of these clusters, 96 are new observations. These targets have X-ray mass estimates and redshifts that formally place them above the thermal Sunyaev-Zel'dovich effect (tSZ) survey limits from the Atacama Cosmology Telescope (ACT, DR5), yet were not detected.  Compact sources with apparent locations close to ($<104''$) the center of a galaxy cluster can in-fill the tSZ flux decrement, resulting in tSZ surveys missing clusters.  To quantify the number of missing clusters from ACT, we carried out a survey at 90~GHz using MUSTANG2 on the Green Bank Telescope achieving a $5\sigma$ detection limit of 1~mJy in the cluster centers. We detected 11 discrete sources, which when scaled, is slightly lower than our previous tSZ selected sample, M2-ACT (8.0\% vs 9.9\%).  Unlike the M2-ACT sample, the sources in X-ray selected clusters were {significantly more} concentrated close to the cluster centers.  {The sources we found would result in 4.5$\pm1$\% of clusters in a tSZ survey being missed -- a result similar to the estimate in recent ACT results (DR6). However, many of the X-ray selected clusters had lower weak lensing masses. A sub-sample of 78 clusters with weak lensing masses better matched to the mass distribution of M2-ACT was made. The source population in these 78 clusters would indicate that 7$\pm2$\% of clusters could be missed by tSZ surveys with a depth similar to DR5.}  

\end{abstract}


\section{Introduction} \label{sec:intro}

As well as being interesting objects in their own right, galaxy clusters are extremely useful tools for probing cosmology. Cluster counts as a function of redshift have been used to estimate cosmological parameters
\citep[e.g.,][]{SPT_constraints} and can help discern between different models for the nature of dark energy \citep{Allen_et_al_2011}. Initially, the accuracy of this method was limited by the size of available catalogs.  However, as catalogs of galaxy clusters have become larger, systematic and selection effects have become more important.

Optical and X-ray surveys have long been used to detect and study galaxy clusters. More recently, millimeter-wave surveys have been used to find clusters via the thermal Sunyaev-Zel'dovich effect \citep[tSZ see][for a review]{MroczkowskiSZReview}.
The tSZ is a spectral distortion of the Cosmic Microwave background (CMB) due to inverse Compton scattering off hot electrons in the intracluster medium (ICM).  As this is nearly redshift-independent, the tSZ is a powerful tool for detecting galaxy clusters.  Experiments such as the Atacama Cosmology Telescope \citep[ACT;][]{ACT} and the South Pole Telescope \citep[SPT;][]{spt} have discovered many clusters.  The latest catalogs from the SPT contain over 7000 optically confirmed clusters \citep{SPTclusters}, while Data Releases 5 and 6 (DR5, DR6) from ACT contain over 4000 and over 10\,000 respectively \citep{Hilton2020,ACTclusters2026}. In the future, the Simons Observatory is expected to discover an order of magnitude more clusters \citep{SimonsForecastPaper}.  To reach percent level accuracy on cosmological parameters such as $\sigma_8$, careful treatment of systematic effects such as contamination by radio sources is needed.

\subsection{Compact sources in clusters}
To maximize their mapping speeds, tSZ survey telescopes have beam sizes comparable to the angular size of clusters ($1\farcm4$--$2\farcm2$ for ACT).  At these resolutions, the signals from compact sources in cluster fields can blend in with the more extended tSZ signal from a cluster. The sources can be contained within the cluster or be in the foreground or background; however, their effect is the same. As the tSZ signal is negative at the frequencies 
where tSZ surveys are most sensitive (the 90, and 150~GHz atmospheric windows), sources {in the apparent center of a cluster} can fully or partially cancel the measured signal.  

Models of the prevalence of sources exist \cite[e.g.][]{Lin2007,Li2022} and it is possible to marginalize over them and recover unbiased cosmological parameters.  {However, there are disagreements between some measurements of compact source populations at 90~GHz and consequently } how much this might contribute to the scatter in $Y$-$M$ relationships \citep{Coble2007,Gralla2020}.  In addition to contributing to the scatter, a strong source {optically aligned to the center of} a cluster could result in a non-detection by a tSZ survey.  Given the well known correlation between radio sources and clusters, the magnitude of this effect {will} be larger than if radio sources and clusters were uncorrelated.  In the recent DR6 data release it was estimated that, even though attempts were made to mitigate compact source contamination, 2-3\% of clusters were missed due to compact sources. Furthermore, up to 7\% may have inaccurate mass estimates caused by miss-centering of the  cluster due to point sources at greater angular separation \citep{ActClusters2025}.

\begin{figure}
    \centering
   \includegraphics[width=0.95\linewidth]{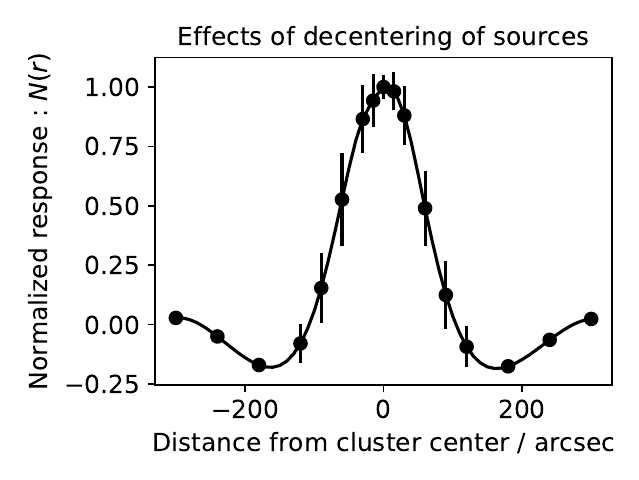}
   \includegraphics[width=0.95\linewidth]{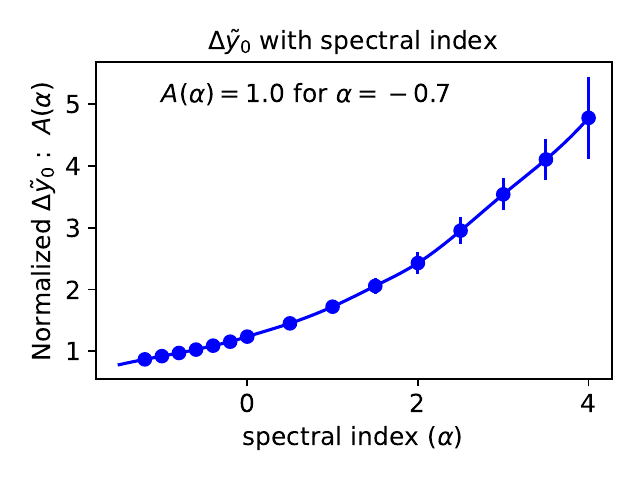}
\caption{How a source of fixed amplitude at 90~GHz affects the recovered Compton-$y$ parameter varies depending on its projected distance from the cluster center and the source spectral index.  [Reproduced from \citet{Dicker2021}].  This can be represented by the functions
   $N(r)$ (upper panel) which has a null at 104$''$ and $A(\alpha)$ (lower panel). The exact shape depends on the matched filters used by a given tSZ survey but will be similar to these examples from ACT DR5. }
              \label{fig:flux2yc}\end{figure}

The magnitude of the effect depends on the flux of the source at each tSZ band (90 and 150~GHz) which can also be expressed as a flux at 90~GHz and a spectral index $\alpha$ between the two.  These spectral indices can vary widely depending on source type, and at 90~GHz are often different from spectral indices calculated at lower frequencies. This makes predictions from fluxes at lower frequencies extremely unreliable, {motivating surveys for sources at the frequencies of interest. In \citet{Dicker2021} ([hereafter D2021]) simulations that added sources to the real ACT maps before extracting clusters with the same software used in DR5 \citep[Nemo:][]{nemo} were carried out. The} effect of a source with a 90~GHz flux $I$ on the measured centralized ACT Compton-$y$ ($y_c$) was found to be:
\begin{equation}\label{equ:dy0}
\Delta_{\tilde{y}_0} = I \:\delta_{\tilde{y}_0}\: N(r)\: A(\alpha)
\end{equation}
where $\delta_{\tilde{y}_0}= -8.76{\times}10^{-6}$ is a constant equal to the effect on the Compton-$y$ of a 1~mJy source at 90 GHz, $r=0$, and with a spectral index of $-0.7$,  and $N(r)$ and $A(\alpha)$ are functions representing the response to a source as a function of projected radius $r$ from the cluster center and spectral index $\alpha$ (Figure~\ref{fig:flux2yc}). {At values of $r<104''$ $N(r)$ is positive meaning that a source decreases the measured tSZ signal, while above $r=104''$, a source makes the cluster signal stronger.} {Although this was specific to DR5, any tSZ survey using multi-frequency matched filters to find clusters will have a similar result.}

\subsection{Past results}

In \citet[hereafter D2024]{Dicker_2024} we presented the results of a blind survey for sources in 248 tSZ selected clusters from the DR5 cluster sample. In that survey, referred to as M2-ACT, only 24 sources were found, significantly fewer than the $70\pm30$ sources predicted from the ad-hoc search for sources from archive data described in \citet{Dicker2021}.  {As might be expected from Equation~\ref{equ:dy0},} compared to the Websky model for compact sources in clusters \citep{Li2022}, fewer bright sources ($>$5 mJy) were found in the centers.  To investigate if this is a selection effect of our tSZ selected sample and to place better limits on the estimates from \citet{ActClusters2025}, here we present the results of a complementary blind survey of X-ray selected clusters.

Our sample selection, observations, and data reduction are in Section~\ref{sec:data}.  Section~\ref{sec:counterparts} describes our search for counterparts at other wavelengths to the sources we found and in Section~\ref{sec:comp} we compare our results with with similar data from our tSZ selected sample and the Websky model \citep{Stein2020}. How the $Y$-$M$ relationship for ACT DR5 is affected by compact sources is covered in Section~\ref{sec:YM} and our conclusions can be found in Section~\ref{sec:conclusions}. We report all errors at a $1\sigma$ / 68.5\% confidence level. We assume the same cosmology as DR5 /DR6, namely a flat $\Lambda$CDM cosmology with $\Omega_m = 0.3$, $\Omega_\Lambda = 0.7$, and $H0 = 70~\mbox{km}\,\mbox{s}^{-1}\,\mbox{Mpc}^{-1}$

\section{Observations} \label{sec:data}
\subsection{Sample selection}
The tSZ selected sample (M2-ACT) in D2024 was constructed by taking all DR5 galaxy clusters in a fixed area of sky.  This area encompasses the eFEDS survey footprint.  eFEDS was the first area of the sky surveyed to the planned survey depth by the eROSITA X-ray satellite \citep{Predehl_2021}. In the approximately 140 square degree area, 542 galaxy clusters and groups were found \citep{Lui2022}.  Many of these have masses and or  redshifts below the sensitivity or completeness limits of the DR5 survey \citep[][Figure 7]{Hilton2020}.  However, a significant number (96) have claimed X-ray derived masses and redshifts that, if true, are likely to ($\mbox{M}_{500c}>2\times10^{14}$, $z>0.175$) or highly likely to ($\mbox{M}_{500c}>4\times10^{14}$, $z>0.175$) have been detected -- yet they were not.  {Here we define likely as coming from a region of mass-redshift space with a median completeness in DR5 of 70\% and highly likely as greater than 96\%.}
In addition to these 96 clusters a further 42 eFEDS clusters were co-detected by ACT DR5 and observed as part of the M2-ACT sample.  Combined, these form an X-ray selected sample of 138 clusters to compare to M2-ACT which we refer to as M2-eFEDS. 

{X-ray catalogs have very different selection effects to tSZE catalogs such as DR5. Important for this study is that compact sources such as AGN always add to the X-ray signal so clusters will not be missed due to the presence of a source. 
Consequently, the number of radio sources found in an X-ray selected sample of clusters can help answer the question of how many clusters might be missed in a tSZE cluster survey.   Other differences are that X-ray catalogs suffer from redshift dimming, while a tSZE survey is less sensitive to nearby clusters due to their large angular extent. For this reason we use the $z>0.175$ cutoff when constructing our sample.  

The eFEDS catalog contains some AGN misidentified as clusters.  \citet{Lui2022} distinguish between compact AGN and larger clusters using an extent likelihood, a measure of how extended the source is. At extent likelihood values greater than 6, up to 20\% of their catalog may not be clusters but at greater than 12 the fraction of misidentified AGN drops to 10\%. The {median} likelihood of the M2-eFEDS clusters is even higher (22) {and only 8\% of the sample have values under 12.} Consequently, we claim that the vast majority ($>$95\%) of the M2-eFEDS sample are likely to be real clusters. }

\subsection{Observations and data reduction}
The 96 unobserved clusters in M2-eFEDS were surveyed at 90~GHz using MUSTANG2 on the Green Bank Telescope (GBT).  To allow easy comparison to D2024 we used the same on-the-fly mapping technique, namely {\em daisy} scans centered on the cluster with a radius of $3'$. With 8 minutes of integration, the noise at the center of the maps is better than 0.14~mJy/beam so that all sources that could affect the measured tSZ signal of a typical $2.6\times10^{14}~\mbox{M}_\odot$ cluster by more than 1\% would be detected at $>5\sigma$.  

\begin{table}
\caption{A partial list of the M2-eFEDS clusters and the depth of each map. Where available, weak lensing masses from \citet{wl_eFEDS} are given. The median WL mass of the full sample is lower than the X-ray derived mass by a factor of 1.9. For those missing from DR5, tSZ masses are found by forced photometry on the ACT DR5 data and make use of the same weak lensing scaling relationship used in DR5. tSZ masses of 0.0 are given for negative Compton-$y$ values { and for clusters with sources, a second tSZ mass with the source removed is given.}   The full table, (with complete names), can be found 
in the machine readable format in the online Journal,
the appendix,
and individual maps can be found on the Harvard Dataverse: \dataset[doi:10.7910/DVN/OXXQBM]{https://doi.org/10.7910/DVN/OXXQBM}. 
}\label{tab:clusters}
\hspace{-1.3cm}\begin{tabular}{cclc} \hline\hline
    Name & Map Noise & X-ray/WL/tSZ mass & Redshift \\
        &  (mJy/beam)     & ($\times10^{14}M_\odot$) & \\\hline
J082820.5 & 0.10 & 6.78 / 3.52 / 2.41 & 0.84 \\
J082955.4 & 0.10 & 4.89 / 1.89 / 2.63 & 0.94 \\
J083204.4 & 0.09 & 2.56 / 1.57 / 1.37,2.13 & 0.20 \\
J083250.0 & 0.11 & 2.59 / 1.03 / 1.72 & 0.80 \\
J083310.3 & 0.12 & 2.25 / 0.94 / 0.21 & 0.75 \\
J083509.0 & 0.13 & 4.93 / 2.47 / 1.53 & 0.60 \\
J083626.4 & 0.12 & 2.71 / NA / 0.12 & 0.57 \\
\ldots    & \ldots & \ldots  & \ldots \\ \hline
\end{tabular}\end{table}

Due to efficient scheduling, we were able to observe all of our eFEDS targets at least once and re-observe bad observations.  The same data reduction pipeline from D2024 was used to make maps and extract sources.  For absolute calibration, we used observations of ALMA calibration sources\footnote{https://almascience.nrao.edu/alma-data/calibrator-catalogue}.  During observing sessions, at least one of these calibrators was observed every 20 minutes.  The calibrated timestreams were turned into maps using the MUSTANG2 filter-and-bin mapmaker \citep[MIDAS; see][for details]{Romero2020}.  The same mapmaker settings, such as the high-pass filtering frequency (0.05~Hz) and map pixel size ($2''$), from D2024 were used.  For these short observations, cutting out all detector glitches is critical -- a few marginal detectors can greatly increase map noise.  Careful manual inspection of the resulting maps was carried out.  In addition to the signal map, a signal free noise map was made by flipping the sign of half of the data from each cluster.  Due to different weather, the noise in each cluster's maps varies (see Table~\ref{tab:clusters}) but the medium noise in the central $1'$ of the maps was 0.15~mJy/beam, 15\% lower than the D2024 sample. 

For maximum sensitivity to compact sources, a matched filter to the MUSTANG2 beam would be used. However, the noise in the MUSTANG2 maps is complex, making an optimal matched filter difficult to calculate.  As with D2024, a good approximation can be made using a difference-of-Gaussians (DoG) filter with inner and outer widths of ($\sigma_1=3\farcs6$ and $\sigma_2=13\farcs6$), respectively. Both the signal and noise maps for each cluster were filtered.    By looking at how the noise varies in the filtered noise maps, signal-to-noise ratio (SNR) maps for each cluster were produced.

\begin{figure}
    \centering
\includegraphics[width=\linewidth]{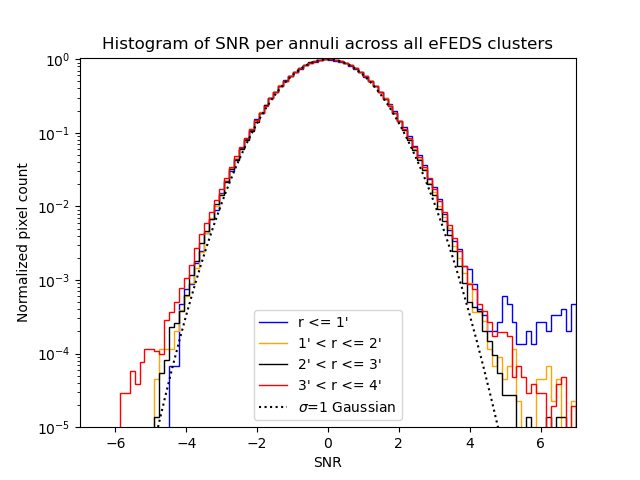}
    \caption{Histograms of the signal-to-noise ratios of  pixels in the MUSTANG2 maps.  Inside a radius of $3'$ the histograms fall off rapidly and a $5\sigma$ cut should produce less than one false positive in our survey which has $\mathcal{O}10^5$ independent beams.  Outside this radius, there are fewer observations per pixel 
and the noise statistics have a significant non-Gaussian tail. Consequently, a $7\sigma$ cut was needed to obtain the same false-positive rate. The excess power on the positive side is due to compact sources.}\label{fig:noise}
\end{figure}

The signal from sources will always be positive. For our target survey depth, the negative signal from the tSZ is negligible and is on angular scales removed by the DoG filter. This allows sources to be found by looking for positive peaks in the SNR maps.   Once a source location is established, a fit for its amplitude and width is carried out in the unfiltered signal map using a 2D Gaussian.  By using the original signal map for this fit, there was no need to account for the effects of the DoG filter.
As with D2024, in the central $r=3'$ of the maps, it was found that the noise is Gaussian and a cut of $5\sigma$ was sufficient to ensure that the chance of a false positive in the whole sample was less than 10\%.  Between $r=3'$ and $r=4'$ the noise has a non-Gaussian tail which we attribute to some map pixels having a lower number of observations.  In this region, a SNR cut of $7\sigma$ was needed to obtain the same false-positive rate  (see Figure~\ref{fig:noise}).  As compact sources at $r>3'$ have less than half the effect on a cluster than an identical source in the center, the higher SNR cut does not compromise our goal of finding all sources that can significantly affect the measured tSZ signal.

\subsection{Sources found}
\begin{figure}
    \centering   
    \includegraphics[width=\linewidth]{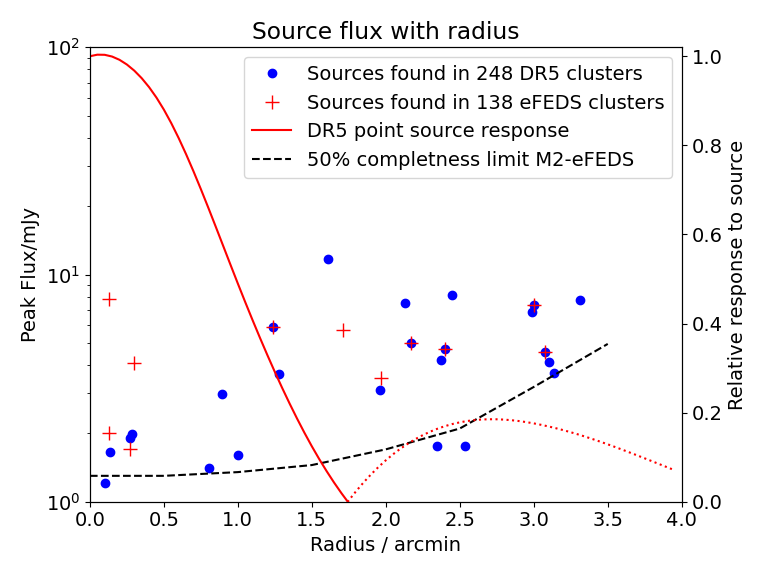}
\caption{The fluxes and distances from the cluster centers of sources found in this survey compared to our previous tSZ selected sample M2-ACT.  Shown in red is the response of the ACT DR5 cluster pipeline to a compact source.  Solid parts of this line are regions where a source reduces the measured Compton-$y$ while in the dashed region a source increases the measured value.  The black dashed line represents the 50\% completeness limit of the M2-eFEDS survey.  Sources from clusters that appeared in both the DR5 and eFEDS catalogs are plotted as a red cross on a blue dot. }\label{fig:sources_found}
\end{figure}

Overall, the M2-eFEDS sample contained 11 sources, 6 in the new observations and 5 from those clusters which were co-detections with DR5 and included in the M2-ACT sample.  Their fluxes and distances from the eFEDs cluster centers are given in Table~\ref{tab:sources} and graphically in Figure~\ref{fig:sources_found}.  The host M2-eFEDS clusters for the sources were generally lower redshift with a median redshift of 0.34 compared to the overall sample of 0.55.  However, there was no clear trend with mass, possibly in part due to the small mass range chosen for the M2-eFEDS sample.

Given that the eFEDs cluster catalog is not 100\% pure we carried out a manual check that the 6 new sources were not false cluster detections.  Aside from eFEDS J092619.9+021208, all the host clusters have weak lensing masses $>1\times 10^{14}~\mbox{M}_\odot$.  Meanwhile eFEDS J092619.9+021208 has an extended likelihood of 23 and is listed in the ACT DR6 catalog with a mass of $2.17\times10^{13}~\mbox{M}_\odot$ ({$>3.5\times10^{14}~\mbox{M}_\odot$ when corrected for the effects of the source}). Thus we conclude all of our sources are either hosted in or along the line of sight to a cluster or group.

\newcommand{\figfourcaption}{Signal to noise plots of our sources and their corresponding spectral energy densities (SEDs). On the SNR plot, the MUSTANG2 source is shown as the red circle along with its detected SNR value.  Sources from other point-source catalogs are shown as the same symbols used on the SED plots. Although error bars are plotted on the SED plot, in many cases, they are not visible.  When a point-source catalog did not contain a counterpart, the approximate detection threshold is shown as a triangle with a one-sided error bar. The fitted radio spectral index between MUSTANG2 and FIRST (or the median spectral index from D2024 when no FIRST counterpart was found) is shown as the blue dotted line. The green dashed line shows a typical 40~K dust spectrum that goes through the MUSTANG2 point.  The SPIRE detections or upper limits are mostly below the dust spectrum curves, indicating that MUSTANG2 sources are dominated by radio emission.   The WISE and 2MASS data indicate the presence of hot gas in some sources, but this has a negligible contribution at 90~GHz. }

\begin{figure}
    \centering
\includegraphics[width=0.95\linewidth]{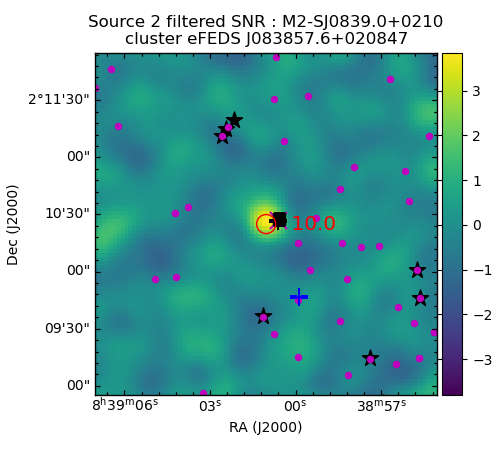}
    \includegraphics[width=0.95\linewidth]{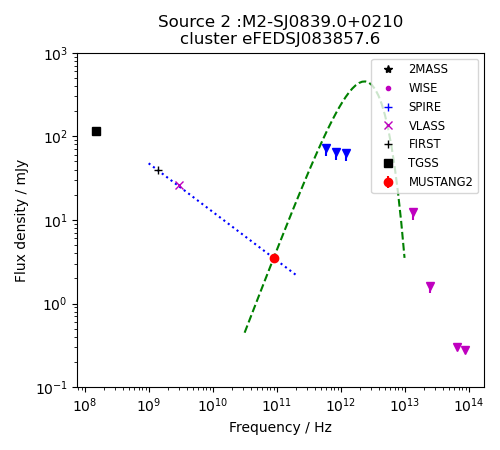}
    \caption{An SNR plot of one of our sources (top) and the corresponding SED plot (bottom).  On the SNR plot, the MUSTANG2 source is shown as the red circle along with its detected SNR value.  Sources from other point-source catalogs are shown as the same symbols used on the SED plots. Although error bars are plotted on the SED plot, in most cases, they are not visible.  When a point-source catalog did not contain a counterpart, the approximate detection threshold is shown as a triangle with a one-sided error bar. The fitted radio spectral index between MUSTANG2 and FIRST (or the median spectral index from D2024 when no FIRST counterpart was found) is shown as the blue dotted line. The green dashed line shows a typical 40~K dust spectrum that goes through the MUSTANG2 point.  The SPIRE detections or upper limits are mostly below the dust spectrum curves, indicating that MUSTANG2 sources are dominated by radio emission.   The WISE and 2MASS data indicate the presence of hot gas in some sources, but this has a negligible contribution at 90~GHz.  
    %The complete figure set (11 images) is available in the online journal. %ApJ
    The complete set of figures can be viewed in the Appendix %astroph
    }\label{fig:examplemap}
\end{figure}

\section{Source Counterparts}\label{sec:counterparts}

As with D2024, searches for counterparts to the sources found by MUSTANG2 were carried out.  Radio catalogs searched included the 150~MHz TGSS \citep{TGSS}, the 1.4~GHz FIRST survey \citep{FIRST}, and the 2--4~GHz VLASS survey \citep{VLASS}.  Higher frequency counterparts were found from SPIRE \citep[600--1200~GHz:][]{SPIRE}, WISE \citep[12--89~THz:][]{WISE_pntsrc}, and the Two Micron All Sky Survey \citep[2MASS, 138--240~THz:][]{2MASS_catalog}. 

We employed a criterion that a matched source is separated by less than the larger FWHM of either MUSTANG2 ($9''$) or the source catalog in question. If no counterpart was found, upper limits from the survey's depth were calculated. The matches found and, where appropriate, the upper limits are shown in the SED plots in the extended version of Figure~\ref{fig:examplemap} and summarized in Table~\ref{tab:sources}.  

Overall, the properties of the sources found in the M2-eFEDS X-ray selected sample were similar to those in the tSZ selected M2-ACT sample.
All M2-eFEDS sources had a counterpart in at least one of the radio catalogs (TGSS, FIRST, and VLASS). However, in many cases, one or more of the other catalogs only contained upper limits.    Spectral indices in the radio, calculated using available flux densities between 1.4 (FIRST) and 90 GHz, were highly varied (-0.85 to 0.37 see Table~\ref{tab:sources}) and in many cases were a poor fit to available data from TGSS and VLASS.  This could be due to either source variability with time or an index that varies with frequency but serves to illustrate the limitations of extrapolating source counts from lower frequencies.  

Matches with WISE are common (8 sources). However, due to the density of WISE sources, {approximately half} of these may be chance alignments.  SPIRE data were available on 5 of the clusters but were all upper limits.  2MASS data were available for 9 clusters, only 3 of which were upper limits. 90~GHz is a regime in which both synchrotron and thermal {dust} emission can be important.     On the assumption that submillimeter and infra red emission from galaxies is well described by a grey body, the high frequency counterparts (or their upper limits) seem to rule out a significant hot dust component in the M2-eFEDS sources.  Fits to a 40~K grey body are shown in SED plots, an example of which is in Figure~\ref{fig:examplemap}.

\begin{table*}
    \caption{Sources found by our survey.  None of the sources were strong enough to enable fitting for extended emission, so only peak fluxes are quoted. A machine readable version of this table that includes the fluxes of counterparts can be found as supplementary material. {Sources marked with $^+$ are clusters with weak lensing masses under $1.5\times10^{14}\mbox{M}_\odot$ and were dropped from the analysis in Section~\ref{sec:sz_comp}}
    }\label{tab:sources}
\hspace{-1.3cm} \scriptsize
\begin{tabular}{llcccccl} \hline 
                  &    & RA & Dec      & Flux   & radius from  &  index & \\
        source ID & eFEDS cluster & J2000 & J2000 & (mJy)  & center & 1.4--90~GHz & Counterparts \\\hline
M2-SJ0832.1$+$0419 & eFEDS J083204.4+041908 & 08h32m04.7s & $+$04$^\circ$18$'$40$''$ & 1.70$\pm$0.20 & 0.27$'$ & -0.83  & TGSS,FIRST,VLASS
\\&&&&&&&,WISE,2MASS\\
M2-SJ0839.0$+$0210 & eFEDS J083857.6+020847 & 08h39m01.0s & $+$02$^\circ$10$'$25$''$ & 3.50$\pm$0.30 & 1.97$'$ & -0.58  & TGSS,FIRST,VLASS\\
M2-SJ0909.6$+$0339 & eFEDS J090930.6+034056 & 09h09m33.4s & $+$03$^\circ$39$'$13$''$ & 5.70$\pm$0.20 & 1.71$'$ & 0.18  &  FIRST,VLASS,WISE
\\&&&&&&&,2MASS\\
M2-SJ0914.6$+$0227$^+$ & eFEDS J091433.9+022718 & 09h14m34.4s & $+$02$^\circ$27$'$29$''$ & 2.00$\pm$0.20 & 0.13$'$ & -0.85  & TGSS,FIRST,VLASS\\
M2-SJ0926.3$+$0212 & eFEDS J092619.9+021208 & 09h26m19.8s & $+$02$^\circ$12$'$27$''$ & 7.80$\pm$0.40 & 0.13$'$ & -0.11  & TGSS,FIRST,VLASS
\\&&&&&&&,WISE\\
M2-SJ0929.4$+$0401 & eFEDS J092921.8+040040 & 09h29m21.9s & $+$04$^\circ$00$'$52$''$ & 4.10$\pm$0.20 & 0.30$'$ & -0.46  & TGSS\\
M2-SJ0839.7$-$0153 & eFEDS J083929.7$-$015005$^*$ & 08h39m40.3s & $-$01$^\circ$52$'$33$''$ & 7.14$\pm$0.34 & 3.00$'$ & 0.28  &  FIRST,VLASS,WISE\\
M2-SJ0839.7$-$0150 & eFEDS J083929.7$-$015005$^*$ & 08h39m43.4s & $-$01$^\circ$50$'$19$''$ & 4.21$\pm$0.38 & 3.08$'$ & 0.37  &  FIRST,WISE,2MASS\\
M2-SJ0901.4$+$0259 & eFEDS J090131.2+030057$^*$ & 09h01m26.1s & $+$02$^\circ$59$'$26$''$ & 4.55$\pm$0.37 & 2.40$'$ & -0.14  &  FIRST,VLASS,WISE
\\&&&&&&&,2MASS\\
M2-SJ0905.7$+$0434 & eFEDS J090540.1+043441$^*$ & 09h05m43.9s & $+$04$^\circ$33$'$42$''$ & 5.80$\pm$0.23 & 1.24$'$ & -0.09  & TGSS,FIRST,VLASS
\\&&&&&&&,WISE,2MASS\\
M2-SJ0918.8$+$0213$^+$ & eFEDS J091849.0+021205$^*$ & 09h18m48.8s & $+$02$^\circ$13$'$26$''$ & 4.87$\pm$0.21 & 2.17$'$ & -0.03  &  FIRST,VLASS,WISE
\\&&&&&&&,2MASS\\
\end{tabular}
\end{table*}

\section{Comparisons with Other cluster samples}\label{sec:comp}
\begin{figure}
    \centering
\includegraphics[width=\linewidth]{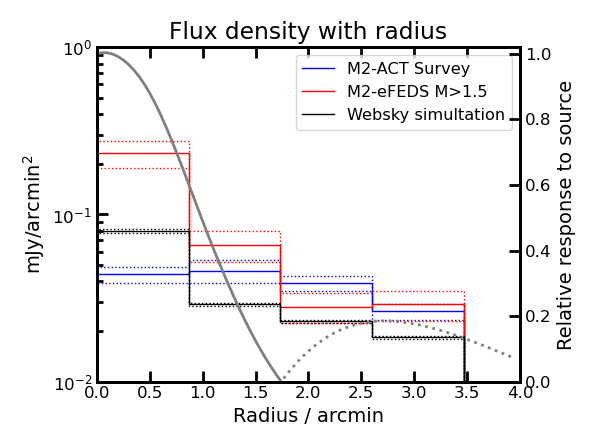}
    \caption{The average 90~GHz flux from point sources as a function of radius for M2-eFEDS {with a $1.5\times 10^{14}\mbox{M}_\odot$ mass cut applied}, M2-ACT from D2024, and the Websky simulation. Plotted on top is the response function of the matched filter used to find the tSZ clusters with the dashed line representing a negative response to a point source. The X-ray sample shows a $4\sigma$ excess of point sources close to the center and a small excess over Websky further out. {If these sources also emit in the X-ray then they are likely to increase the chance of cluster detection in X-ray surveys.}  Conversely, the tSZ selected sample shows an excess only around a radius of  $2\farcm5$ where sources will increase the tSZ signal. Dotted lines represent estimated $1\sigma$\ errors.}
    \label{fig:flux_w_radius}
\end{figure}

\subsection{Comparisons with tSZ selected samples}\label{sec:sz_comp}
{In the complete samples, there is no statistically significant difference in the prevalence of sources between the M2-ACT and M2-eFEDS samples of clusters (9.9\% vs. 8.0\%). However, from Table~\ref{tab:clusters}, it can be seen that the weak lensing (WL) masses \citep{wl_eFEDS} of the eFEDS sample are considerably lower than the X-ray masses with a median ratio of 1.9.  If one accepts the WL masses as the true values then, fractionally, there are more low mass clusters in the M2-eFEDS sample than M2-ACT.  As they have intrinsically fewer galaxies and lower feedback onto their central galaxy, lower mass clusters are less likely to have radio sources \citep{src_vs_clustermass}. A two-sided Kolmogorov–Smirnov (KS) test to see if the masses of the full samples could come from the same parent population fails with a $p$-value of $<0.1\%$.  However, if one cuts low mass clusters from the M2-eFEDS sample, the distribution between WL masses in M2-eFEDS and tSZ masses in M2-ACT becomes similar with a broad maximum in the $p$-value of 66\% at a mass cut of $1.5\times10^{14} \, \mbox{M}_\odot$.  For eFEDS clusters without WL masses we used the X-ray mass scaled down by 1.9. 
The resulting subsample of 78 clusters from the M2-eFEDS sample has a much better match to the M2-ACT mass distribution and a total source prevalence in the sub-sample of 11.5\%.

A plot of the average flux in radial bins of this subsample can be seen in Figure\,\ref{fig:flux_w_radius}.  There is a $4\sigma$ excess of flux in the central bin from the X-ray clusters due to centrally placed sources.  Sources in this region suppress the tSZ signal lowering the chance of detection and, if they have X-ray counterparts, increase the chances of a cluster being detected in the X-ray.   Further out, ($>104''$) the distribution of the fluxes becomes similar between the two samples.  In this region, sources increase both the tSZ and the X-ray signals.

\subsection{Comparisons with Websky}\label{sec:websky}
Websky is a set of simulated observations of the extragalactic microwave sky \citep{Stein2020}. Included in this simulation are catalogs of radio sources \citep{Li2022} and galaxy clusters.  The full catalog contains many sources below the sensitivities of our MUSTANG2 observations and clusters at a larger range of redshift and mass.  To obtain a sample to compare against, we first selected a subsample of $\sim$10,000 clusters that matched the same mass and redshift range as M2-ACT.  All Websky sources within $5'$ of each Websky cluster were found and the completeness of the MUSTANG2 observations applied.  This completeness (as a function of flux and radius from the cluster center) was found using simulated sources in real MUSTANG2 data.

The Websky sources are always placed in the center of cluster halos and the only sources further out are those not associated with the cluster.  In the central $0.9'$ of the clusters, Websky matches well the results from M2-ACT while the flux density in M2-eFEDS is significantly higher ($3\sigma)$.  Further out, both M2-eFEDS and M2-ACT are higher than websky.  The overall typical flux from sources in the Websky sample is lower than both measured samples indicating that Websky does not do a good job predicting 90~GHz source counts away from the cluster centers.  If M2-eFEDS represents the true distribution this is also true in the cluster centers.
}

\begin{figure}
    \centering
\includegraphics[width=\linewidth]{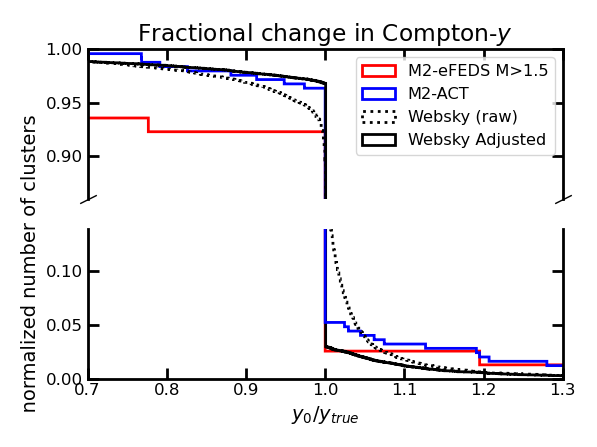}
\caption{Cumulative plots of the fractional change in measured Compton-$y$ caused by sources calculated using 
    Equation~\ref{equ:dy0}.  The red curve is for M2-eFEDS while the blue curve is for the tSZ selected sample, M2-ACT (from D2024). Results using sources from the Websky simulation are in black. The dotted black line shows the results using all sources in the Websky sample above 0.5~mJy {(that would be observed with a very deep survey)}, while the solid line shows the the same sample taking into account the MUSTANG2 survey depths.\label{fig:dy}}
\end{figure}

\section{Implications for the ACT $Y$-$M$ relation}\label{sec:YM}

A more quantitative comparison between the samples can be made using Equation~\ref{equ:dy0} to predict the difference between the measured tSZ signal ($y_0$) from its true value ($y_{\mbox{true}}$). {For correct comparison with DR5 we use the ACT $Y$-$M$ relation, $y_{\mbox{true}}$ is calculated without point sources and $y_0$ includes the effects of point sources on the tSZE measurement. For M2-eFEDS we use the WL masses where available and the scaled X-ray mass when no WL masses are available. }
Figure~\ref{fig:dy} shows a cumulative histogram of the results from D2024 along with the new M2-eFEDS sample {with a mass cut of $1.5\times10^{14}\mbox{M}_\odot$}.  To show the effects of the depth of our surveys, the Websky sample is plotted with and without the completeness of our samples applied.  Without the completeness cut, the Websky sample contains many sources below our {finite} survey depth and shows a more gradual response around  $y_0 / y_{\mbox{true}}=1$.  This is more reflective of reality but is not good for comparison to our observations which were designed to detect sources that would produce more significant ($>1$\%) changes.

{Central sources produce values of $y_0 / y_{\mbox{true}}$ below 1 and cause the largest differences between the M2-ACT and M2-eFEDS samples, with 7$\pm$2\% of clusters in the X-ray selected sample having a measured value more than 25\% lower. If the mass cut on the X-ray sample is not carried out then this number drops to 4.5$\pm$1\%.  This compares to just 1.5\% for the M2-ACT and Websky samples.  The difference here represents clusters that are potentially missed by tSZ surveys and a possible deficit of central sources in Websky.   Given our sample size and that it has lower mass limits, this is consistent with the value in 
\citet{ActClusters2025} (2--3\%). }This result is not highly dependent on the detection threshold used in DR5 either -- due to the cluster mass function, the majority of clusters are detected close to the detection threshold and reductions in the measured tSZ signal by 25\% are enough to reduce most clusters below the survey limit. 
 Sources at radii larger than $104''$ increase the measured tSZ signal. As expected from Figure~\ref{fig:flux_w_radius}, the X-ray selected M2-eFEDS and M2-ACT  samples show more clusters with $y_0 / y_{\mbox{true}}>1$ than websky.
 
\section{Conclusions}\label{sec:conclusions}
Our blind M2-eFEDS survey of 138 X-ray selected clusters contained 11 sources for which the 90 GHz flux density is dominated by radio emission. 
The fraction of these clusters having sources (8.0\%) was similar to the results from the M2-ACT sample (tSZ selected ) in D2024 (9.9\%).  The M2-ACT clusters were selected only by sky area and being included in the ACT DR5 cluster catalog and so the only biases in that sample should be those native to tSZ surveys.  The X-ray sample was selected over the same region of sky. However, the X-ray masses the sample was based off were biased high by a factor of 1.9 when compared to weak lensing masses (see Table~\ref{tab:clusters}).  {If one takes the weak lensing masses as correct then comparison of the complete samples is biased due to the greater number of low mass clusters in M2-eFEDS. A mass cut of $1.5\times10^{14}\, \mbox{M}_\odot$ on the weak lensing masses in M2-eFEDS gives a similar mass distribution to M2-ACT.  This leaves a sample of 78 clusters of which 11.5\% have sources.} Although the M2-ACT and M2-eFEDS samples were selected using DR5, since they were taken, the DR6 ACT catalog has been released.  The lower noise in DR6 resulted in 27 of the 98 clusters seen in eFEDS but not DR5 being detected although at much lower masses than reported in in the X-ray.

Although the prevalence of sources is similar between the two cluster samples, their locations within their host clusters is not.  
The average flux from sources is significantly higher in the centers of the X-ray clusters.  {Outside of a radius of $104''$, both the tSZ and X-ray selected clusters show more sources than the Websky model.  This radius is where the response function of the filter used to find clusters in ACT DR5 changes from lowering the measured tSZ signal to increasing to it. Using the framework from D2021, it was possible to show that the extra sources in the centers of the X-ray sample would result in 5.5\% more of these clusters having measured tSZ signals lower than their true value by 25\% or more.  Such clusters are less likely to be detected by tSZ surveys. The inclusion of more lower mass clusters lowers the fraction of clusters that could be missed.  With the complete M2-eFEDs sample the additional fraction of clusters with a 25\% or more reduction drops to 3\%.  This is consistent with the estimation from the newly released ACT cluster catalog DR6 which includes such lower mass clusters.  A larger sample with better matched masses, would allow  more accurate numbers and better distinguish between the different selection effects in the X-ray and tSZ samples.  However, our observations show that sources will not be a significant problem in upcoming tSZ surveys. }

\section*{acknowledgments}
The GBT data were acquired under the project ID AGBT24B\_298.  The Green Bank Observatory is a facility of the National Science Foundation operated under cooperative agreement by Associated Universities, Inc. This publication makes use of data products from the Wide-field Infrared Survey Explorer, which is a joint project of the University of California, Los Angeles, and the Jet Propulsion Laboratory/California Institute of Technology, funded by the National Aeronautics and Space Administration.
S.R.D. is supported by NSF grant No. 2307546. 
This publication makes use of data products from the Wide-field Infrared Survey Explorer \citep{vo:WISE_scs}, which is a joint project of the University of California, Los Angeles, and the Jet Propulsion Laboratory/California Institute of Technology, funded by the National Aeronautics and Space Administration. {We would like to thank the referee for their helpful feedback.}

\vspace{5mm}
\facilities{GBT (MUSTANG2), WISE, Herschel (SPIRE), eROSITA}

\software{astropy \citep{2013A&A...558A..33A,2018AJ....156..123A}}

\bibliography{new.ms}{}

@ARTICLE{2MASS_catalog,
       author = {{Jarrett}, T.~H. and {Chester}, T. and {Cutri}, R. and {Schneider}, S. and {Skrutskie}, M. and {Huchra}, J.~P.},
        title = "{2MASS Extended Source Catalog: Overview and Algorithms}",
      journal = {\aj},
         year = 2000,
        month = may,
       volume = {119},
       number = {5},
        pages = {2498-2531},
          doi = {10.1086/301330},
archivePrefix = {arXiv},
       eprint = {astro-ph/0004318},
 primaryClass = {astro-ph},
       adsurl = {https://ui.adsabs.harvard.edu/abs/2000AJ....119.2498J}
}

@software{nemo,
       author = {{Hilton}, Matt and {Aiola}, Simone and {Alonso}, David and {Hasselfield}, Matthew and {Huffenberger}, Kevin and {Marriage}, Toby and {MacCrann}, Niall and {Naess}, Sigurd and {Sif{\'o}n}, Crist{\'o}bal},
        title = "{Nemo: Millimeter-wave map filtering and Sunyaev-Zel'dovich galaxy cluster and source detection}",
 howpublished = {Astrophysics Source Code Library, record ascl:2308.006},
         year = 2023,
        month = aug,
          eid = {ascl:2308.006},
archivePrefix = {ascl},
       eprint = {2308.006},
       adsurl = {https://ui.adsabs.harvard.edu/abs/2023ascl.soft08006H}
}

@ARTICLE{SPTclusters,
       author = {{Bleem}, L.~E. and {Klein}, M. and {Kornoelje}, K. and {Bocquet}, S. and {Sobrin}, J.~A. and {Aguena}, M. and {Anderes}, E. and {Anderson}, A.~J. and {Andrade-Oliveira}, F. and {Ansarinejad}, B. and {Archipley}, M. and {Balkenhol}, L. and {Barron}, D.~R. and {Barry}, P.~S. and {Bayliss}, M. and {Benabed}, K. and {Bender}, A.~N. and {Benson}, B.~A. and {Bianchini}, F. and {Bouchet}, F.~R. and {Brooks}, D. and {Burke}, D.~L. and {Calzadilla}, M. and {Camilleri}, R. and {Camphuis}, E. and {Campitiello}, M.~G. and {Carlstrom}, J.~E. and {Carnero Rosell}, A. and {Carretero}, J. and {Carron}, J. and {Chang}, C.~L. and {Chichura}, P.~M. and {Chokshi}, A. and {Chou}, T.-L. and {Coerver}, A. and {Costanzi}, M. and {Crawford}, T.~M. and {da Costa}, L.~N. and {Daley}, C. and {Davis}, T.~M. and {de Haan}, T. and {De Vicente}, J. and {Desai}, S. and {Dibert}, K.~R. and {Diehl}, H.~T. and {Dobbs}, M.~A. and {Doohan}, M. and {Dutcher}, D. and {Everett}, S. and {Evrard}, G. and {Feng}, C. and {Ferguson}, K.~R. and {Ferree}, N.~C. and {Fichman}, K. and {Flaugher}, B. and {Floyd}, B. and {Fosdick}, K. and {Foster}, A. and {Galli}, S. and {Gambrel}, A.~E. and {Gao}, A.~K. and {Garc{\'\i}a-Bellido}, J. and {Ge}, F. and {Gladders}, M.~D. and {Grandis}, S. and {Guidi}, F. and {Guns}, S. and {Gutierrez}, G. and {Halverson}, N.~W. and {Hinton}, S.~R. and {Hivon}, E. and {Holder}, G.~P. and {Hollowood}, D.~L. and {Holzapfel}, W.~L. and {Hood}, J.~C. and {Hryciuk}, A. and {Huang}, N. and {Jhaveri}, T. and {Kent}, S. and {K{\'e}ruzor{\'e}}, F. and {Khalife}, A.~R. and {Khullar}, G. and {Knox}, L. and {Kuehn}, K. and {Kuo}, C.-L. and {Lahav}, O. and {Levy}, K. and {Li}, Y. and {Lowitz}, A.~E. and {Lu}, C. and {Lynch}, G.~P. and {Maccarone}, T.~J. and {Mahler}, G. and {Maniyar}, A.~S. and {Marshall}, J.~L. and {Martsen}, E.~S. and {McDonald}, M. and {Mena-Fern{\'a}ndez}, J. and {Menanteau}, F. and {Millea}, M. and {Miquel}, R. and {Mohr}, J.~J. and {Montgomery}, J. and {Myles}, J. and {Nakato}, Y. and {Natoli}, T. and {Nichol}, R.~C. and {Novosad}, V. and {Ogando}, R.~L.~C. and {Omori}, Y. and {Ouellette}, A. and {Pan}, Z. and {Phadke}, K.~A. and {Plazas Malag{\'o}n}, A.~A. and {Pollak}, A.~W. and {Porredon}, A. and {Prabhu}, K. and {Prat}, J. and {Quan}, W. and {Raghunathan}, S. and {Rahimi}, M. and {Rahlin}, A. and {Reichardt}, C.~L. and {Romer}, A.~K. and {Rouble}, M. and {Ruhl}, J.~E. and {Sanchez}, E. and {Sanchez Cid}, D. and {Schrabback}, T. and {Sevilla-Noarbe}, I. and {Silva Oliveira}, A.~C. and {Simpson}, A. and {Somboonpanyakul}, T. and {Stark}, A.~A. and {Suchyta}, E. and {Swanson}, M.~E.~C. and {Tandoi}, C. and {To}, C. and {Trendafilova}, C. and {Vieira}, J.~D. and {Vieregg}, A.~G. and {Vikram}, V. and {Vitrier}, A. and {Wan}, Y. and {Weaverdyck}, N. and {Weller}, J. and {Whitehorn}, N. and {Wu}, W.~L.~K. and {Young}, M.~R. and {Zebrowski}, J.~A. and {Zhan}, J.},
        title = "{Galaxy Clusters Selected via the Sunyaev-Zel'dovich Effect in 5 year data from the SPT-3G Main Survey}",
      journal = {arXiv e-prints},
         year = 2026,
        month = jul,
          eid = {arXiv:2607.01175},
        pages = {arXiv:2607.01175},
          doi = {10.48550/arXiv.2607.01175},
archivePrefix = {arXiv},
       eprint = {2607.01175},
 primaryClass = {astro-ph.CO},
       adsurl = {https://ui.adsabs.harvard.edu/abs/2026arXiv260701175B}
}

@ARTICLE{spt,
       author = {{Carlstrom}, J.~E. and {Ade}, P.~A.~R. and {Aird}, K.~A. and {Benson}, B.~A. and {Bleem}, L.~E. and {Busetti}, S. and {Chang}, C.~L. and {Chauvin}, E. and {Cho}, H.-M. and {Crawford}, T.~M. and {Crites}, A.~T. and {Dobbs}, M.~A. and {Halverson}, N.~W. and {Heimsath}, S. and {Holzapfel}, W.~L. and {Hrubes}, J.~D. and {Joy}, M. and {Keisler}, R. and {Lanting}, T.~M. and {Lee}, A.~T. and {Leitch}, E.~M. and {Leong}, J. and {Lu}, W. and {Lueker}, M. and {Luong-Van}, D. and {McMahon}, J.~J. and {Mehl}, J. and {Meyer}, S.~S. and {Mohr}, J.~J. and {Montroy}, T.~E. and {Padin}, S. and {Plagge}, T. and {Pryke}, C. and {Ruhl}, J.~E. and {Schaffer}, K.~K. and {Schwan}, D. and {Shirokoff}, E. and {Spieler}, H.~G. and {Staniszewski}, Z. and {Stark}, A.~A. and {Tucker}, C. and {Vanderlinde}, K. and {Vieira}, J.~D. and {Williamson}, R.},
        title = "{The 10 Meter South Pole Telescope}",
      journal = {\pasp},
         year = 2011,
        month = may,
       volume = {123},
       number = {903},
        pages = {568},
          doi = {10.1086/659879},
archivePrefix = {arXiv},
       eprint = {0907.4445},
 primaryClass = {astro-ph.IM},
       adsurl = {https://ui.adsabs.harvard.edu/abs/2011PASP..123..568C}
}

@ARTICLE{SPT_constraints,
       author = {{Bocquet}, S. and {Grandis}, S. and {Bleem}, L.~E. and {Klein}, M. and {Mohr}, J.~J. and {Schrabback}, T. and {Abbott}, T.~M.~C. and {Ade}, P.~A.~R. and {Aguena}, M. and {Alarcon}, A. and {Allam}, S. and {Allen}, S.~W. and {Alves}, O. and {Amon}, A. and {Anderson}, A.~J. and {Annis}, J. and {Ansarinejad}, B. and {Austermann}, J.~E. and {Avila}, S. and {Bacon}, D. and {Bayliss}, M. and {Beall}, J.~A. and {Bechtol}, K. and {Becker}, M.~R. and {Bender}, A.~N. and {Benson}, B.~A. and {Bernstein}, G.~M. and {Bhargava}, S. and {Bianchini}, F. and {Brodwin}, M. and {Brooks}, D. and {Bryant}, L. and {Campos}, A. and {Canning}, R.~E.~A. and {Carlstrom}, J.~E. and {Carnero Rosell}, A. and {Carrasco Kind}, M. and {Carretero}, J. and {Castander}, F.~J. and {Cawthon}, R. and {Chang}, C.~L. and {Chang}, C. and {Chaubal}, P. and {Chen}, R. and {Chiang}, H.~C. and {Choi}, A. and {Chou}, T.-L. and {Citron}, R. and {Corbett Moran}, C. and {Cordero}, J. and {Costanzi}, M. and {Crawford}, T.~M. and {Crites}, A.~T. and {da Costa}, L.~N. and {Pereira}, M.~E.~S. and {Davis}, C. and {Davis}, T.~M. and {DeRose}, J. and {Desai}, S. and {de Haan}, T. and {Diehl}, H.~T. and {Dobbs}, M.~A. and {Dodelson}, S. and {Doux}, C. and {Drlica-Wagner}, A. and {Eckert}, K. and {Elvin-Poole}, J. and {Everett}, S. and {Everett}, W. and {Ferrero}, I. and {Fert{\'e}}, A. and {Flores}, A.~M. and {Frieman}, J. and {Gallicchio}, J. and {Garc{\'\i}a-Bellido}, J. and {Gatti}, M. and {George}, E.~M. and {Giannini}, G. and {Gladders}, M.~D. and {Gruen}, D. and {Gruendl}, R.~A. and {Gupta}, N. and {Gutierrez}, G. and {Halverson}, N.~W. and {Harrison}, I. and {Hartley}, W.~G. and {Herner}, K. and {Hinton}, S.~R. and {Holder}, G.~P. and {Hollowood}, D.~L. and {Holzapfel}, W.~L. and {Honscheid}, K. and {Hrubes}, J.~D. and {Huang}, N. and {Hubmayr}, J. and {Huff}, E.~M. and {Huterer}, D. and {Irwin}, K.~D. and {James}, D.~J. and {Jarvis}, M. and {Khullar}, G. and {Kim}, K. and {Knox}, L. and {Kraft}, R. and {Krause}, E. and {Kuehn}, K. and {Kuropatkin}, N. and {K{\'e}ruzor{\'e}}, F. and {Lahav}, O. and {Lee}, A.~T. and {Leget}, P.-F. and {Li}, D. and {Lin}, H. and {Lowitz}, A. and {MacCrann}, N. and {Mahler}, G. and {Mantz}, A. and {Marshall}, J.~L. and {McCullough}, J. and {McDonald}, M. and {McMahon}, J.~J. and {Mena-Fern{\'a}ndez}, J. and {Menanteau}, F. and {Meyer}, S.~S. and {Miquel}, R. and {Montgomery}, J. and {Myles}, J. and {Natoli}, T. and {Navarro-Alsina}, A. and {Nibarger}, J.~P. and {Noble}, G.~I. and {Novosad}, V. and {Ogando}, R.~L.~C. and {Omori}, Y. and {Padin}, S. and {Pandey}, S. and {Paschos}, P. and {Patil}, S. and {Pieres}, A. and {Plazas Malag{\'o}n}, A.~A. and {Porredon}, A. and {Prat}, J. and {Pryke}, C. and {Raveri}, M. and {Reichardt}, C.~L. and {Roberson}, J. and {Rollins}, R.~P. and {Romero}, C. and {Roodman}, A. and {Ruhl}, J.~E. and {Rykoff}, E.~S. and {Saliwanchik}, B.~R. and {Salvati}, L. and {S{\'a}nchez}, C. and {Sanchez}, E. and {Sanchez Cid}, D. and {Saro}, A. and {Schaffer}, K.~K. and {Secco}, L.~F. and {Sevilla-Noarbe}, I. and {Sharon}, K. and {Sheldon}, E. and {Shin}, T. and {Sievers}, C. and {Smecher}, G. and {Smith}, M. and {Somboonpanyakul}, T. and {Sommer}, M. and {Stalder}, B. and {Stark}, A.~A. and {Stephen}, J. and {Strazzullo}, V. and {Suchyta}, E. and {Tarle}, G. and {To}, C. and {Troxel}, M.~A. and {Tucker}, C. and {Tutusaus}, I. and {Varga}, T.~N. and {Veach}, T. and {Vieira}, J.~D. and {Vikhlinin}, A. and {von der Linden}, A. and {Wang}, G. and {Weaverdyck}, N. and {Weller}, J. and {Whitehorn}, N. and {Wu}, W.~L.~K. and {Yanny}, B. and {Yefremenko}, V. and {Yin}, B. and {Young}, M. and {Zebrowski}, J.~A. and {Zhang}, Y. and {Zohren}, H. and {Zuntz}, J. and {(SPT} and {DES Collaborations)}},
        title = "{SPT clusters with DES and HST weak lensing. II. Cosmological constraints from the abundance of massive halos}",
      journal = {\prd},
         year = 2024,
        month = oct,
       volume = {110},
       number = {8},
          eid = {083510},
        pages = {083510},
          doi = {10.1103/PhysRevD.110.083510},
archivePrefix = {arXiv},
       eprint = {2401.02075},
 primaryClass = {astro-ph.CO},
       adsurl = {https://ui.adsabs.harvard.edu/abs/2024PhRvD.110h3510B}
}

@ARTICLE{ACTclusters2026,
       author = {{Aguena}, M. and {Aiola}, S. and {Allam}, S. and {Andrade-Oliveira}, F. and {Bacon}, D. and {Bahcall}, N. and {Battaglia}, N. and {Battistelli}, E.~S. and {Bocquet}, S. and {Bolliet}, B. and {Bond}, J.~R. and {Brooks}, D. and {Calabrese}, E. and {Carretero}, J. and {Choi}, S.~K. and {da Costa}, L.~N. and {Costanzi}, M. and {Coulton}, W. and {Davis}, T.~M. and {Desai}, S. and {Devlin}, M.~J. and {Dicker}, S. and {Doel}, P. and {Duivenvoorden}, A.~J. and {Dunkley}, J. and {Ferraro}, S. and {Flaugher}, B. and {Frieman}, J. and {Gallardo}, P.~A. and {Gatti}, M. and {Gaztanaga}, E. and {Gill}, A.~S. and {Golec}, J.~E. and {Gruen}, D. and {Gruendl}, R.~A. and {Halpern}, M. and {Hasselfield}, M. and {Hill}, J.~C. and {Hilton}, M. and {Hincks}, A.~D. and {Hinton}, S.~R. and {Hollowood}, D.~L. and {Honscheid}, K. and {Hubmayr}, J. and {Huffenberger}, K.~M. and {Hughes}, J.~P. and {James}, D.~J. and {Klein}, M. and {Knowles}, K. and {Koopman}, B.~J. and {Kosowsky}, A. and {Lahav}, O. and {Lee}, E. and {Lin}, Y. and {Lokken}, M. and {Madhavacheril}, M.~S. and {Malag{\'o}n}, A.~A. Plazas and {Marrewijk}, J. v. and {Marshall}, J.~L. and {McMahon}, J. and {Mena-Fern{\'a}ndez}, J. and {Miquel}, R. and {Miyatake}, H. and {Mohr}, J.~J. and {Moodley}, K. and {Mroczkowski}, T. and {Naess}, S. and {Nati}, F. and {Nicola}, A. and {Niemack}, M.~D. and {Ogando}, R.~L.~C. and {Oguri}, M. and {Orlowski-Scherer}, J. and {Page}, L.~A. and {Partridge}, B. and {da Silva Pereira}, M.~E. and {Porredon}, A. and {Qu}, F.~J. and {Ragavan}, D.~C. and {Guachalla}, B. Ried and {Romer}, A.~K. and {Rosell}, A. Carnero and {Rykoff}, E.~S. and {Samuroff}, S. and {Sanchez}, E. and {Sevilla-Noarbe}, I. and {Sierra}, C. and {Sif{\'o}n}, C. and {Smith}, M. and {Staggs}, S.~T. and {Suchyta}, E. and {Swanson}, M.~E.~C. and {Tucker}, D.~L. and {Vargas}, C. and {Vavagiakis}, E.~M. and {De Vicente}, J. and {Weaverdyck}, N. and {Weller}, J. and {Wollack}, E.~J. and {Zubeldia}, I.},
        title = "{The Atacama Cosmology Telescope: DR6 Sunyaev-Zel'dovich Selected Galaxy Clusters Catalog}",
      journal = {The Open Journal of Astrophysics},
         year = 2026,
        month = jan,
       volume = {9},
        pages = {55863},
          doi = {10.33232/001c.155863},
archivePrefix = {arXiv},
       eprint = {2507.21459},
 primaryClass = {astro-ph.CO},
       adsurl = {https://ui.adsabs.harvard.edu/abs/2026OJAp....955863A}
}

@ARTICLE{ActClusters2025,
       author = {{ACT DES HSC Collaboration} and {Aguena}, M. and {Aiola}, S. and {Allam}, S. and {Andrade-Oliveira}, F. and {Bacon}, D. and {Bahcall}, N. and {Battaglia}, N. and {Battistelli}, E.~S. and {Bocquet}, S. and {Bolliet}, B. and {Bond}, J.~R. and {Brooks}, D. and {Calabrese}, E. and {Carretero}, J. and {Choi}, S.~K. and {da Costa}, L.~N. and {Costanzi}, M. and {Coulton}, W. and {Davis}, T.~M. and {Desai}, S. and {Devlin}, M.~J. and {Dicker}, S. and {Doel}, P. and {Duivenvoorden}, A.~J. and {Dunkley}, J. and {Ferraro}, S. and {Flaugher}, B. and {Frieman}, J. and {Gallardo}, P.~A. and {Gatti}, M. and {Gaztanaga}, E. and {Gill}, A.~S. and {Golec}, J.~E. and {Gruen}, D. and {Gruendl}, R.~A. and {Halpern}, M. and {Hasselfield}, M. and {Hill}, J.~C. and {Hilton}, M. and {Hincks}, A.~D. and {Hinton}, S.~R. and {Hollowood}, D.~L. and {Honscheid}, K. and {Hubmayr}, J. and {Huffenberger}, K.~M. and {Hughes}, J.~P. and {James}, D.~J. and {Klein}, M. and {Knowles}, K. and {Koopman}, B.~J. and {Kosowsky}, A. and {Lahav}, O. and {Lee}, E. and {Lin}, Y. and {Lokken}, M. and {Madhavacheril}, M.~S. and {Plazas Malag{\'o}n}, A.~A. and {Marrewijk}, J. v. and {Marshall}, J.~L. and {McMahon}, J. and {Mena-Fern{\'a}ndez}, J. and {Miquel}, R. and {Miyatake}, H. and {Mohr}, J.~J. and {Moodley}, K. and {Mroczkowski}, T. and {Naess}, S. and {Nati}, F. and {Nicola}, A. and {Niemack}, M.~D. and {Ogando}, R.~L.~C. and {Oguri}, M. and {Orlowski-Scherer}, J. and {Page}, L.~A. and {Partridge}, B. and {da Silva Pereira}, M.~E. and {Porredon}, A. and {Qu}, F.~J. and {Ragavan}, D.~C. and {Ried Guachalla}, B. and {Romer}, A.~K. and {Carnero Rosell}, A. and {Rykoff}, E.~S. and {Samuroff}, S. and {Sanchez}, E. and {Sevilla-Noarbe}, I. and {Sierra}, C. and {Sif{\'o}n}, C. and {Smith}, M. and {Staggs}, S.~T. and {Suchyta}, E. and {Swanson}, M.~E.~C. and {Tucker}, D.~L. and {Vargas}, C. and {Vavagiakis}, E.~M. and {De Vicente}, J. and {Weaverdyck}, N. and {Weller}, J. and {Wollack}, E.~J. and {Zubeldia}, I.},
        title = "{The Atacama Cosmology Telescope: DR6 Sunyaev-Zel'dovich Selected Galaxy Clusters Catalog}",
      journal = {arXiv e-prints},
         year = 2025,
        month = jul,
          eid = {arXiv:2507.21459},
        pages = {arXiv:2507.21459},
          doi = {10.48550/arXiv.2507.21459},
archivePrefix = {arXiv},
       eprint = {2507.21459},
 primaryClass = {astro-ph.CO},
       adsurl = {https://ui.adsabs.harvard.edu/abs/2025arXiv250721459A}
}

@ARTICLE{Lin2007,
       author = {{Lin}, Yen-Ting and {Mohr}, Joseph J.},
        title = "{Radio Sources in Galaxy Clusters: Radial Distribution, and 1.4 GHz and K-band Bivariate Luminosity Function}",
      journal = {\apjs},
         year = 2007,
        month = may,
       volume = {170},
       number = {1},
        pages = {71-94},
          doi = {10.1086/513565},
archivePrefix = {arXiv},
       eprint = {astro-ph/0612521},
 primaryClass = {astro-ph},
       adsurl = {https://ui.adsabs.harvard.edu/abs/2007ApJS..170...71L}
}

@ARTICLE{src_vs_clustermass,
       author = {{Best}, P.~N. and {von der Linden}, A. and {Kauffmann}, G. and {Heckman}, T.~M. and {Kaiser}, C.~R.},
        title = "{On the prevalence of radio-loud active galactic nuclei in brightest cluster galaxies: implications for AGN heating of cooling flows}",
      journal = {\mnras},
         year = 2007,
        month = aug,
       volume = {379},
       number = {3},
        pages = {894-908},
          doi = {10.1111/j.1365-2966.2007.11937.x},
archivePrefix = {arXiv},
       eprint = {astro-ph/0611197},
 primaryClass = {astro-ph},
       adsurl = {https://ui.adsabs.harvard.edu/abs/2007MNRAS.379..894B}
}

@ARTICLE{Allen_et_al_2011,
       author = {{Allen}, Steven W. and {Evrard}, August E. and {Mantz}, Adam B.},
        title = "{Cosmological Parameters from Observations of Galaxy Clusters}",
      journal = {\araa},
         year = 2011,
        month = sep,
       volume = {49},
       number = {1},
        pages = {409-470},
          doi = {10.1146/annurev-astro-081710-102514},
archivePrefix = {arXiv},
       eprint = {1103.4829},
 primaryClass = {astro-ph.CO},
       adsurl = {https://ui.adsabs.harvard.edu/abs/2011ARA&A..49..409A}
}

@ARTICLE{ACT,
       author = {{Thornton}, R.~J. and {Ade}, P.~A.~R. and {Aiola}, S. and
         {Angil{\`e}}, F.~E. and {Amiri}, M. and {Beall}, J.~A. and
         {Becker}, D.~T. and {Cho}, H. -M. and {Choi}, S.~K. and {Corlies}, P. and
         {Coughlin}, K.~P. and {Datta}, R. and {Devlin}, M.~J. and
         {Dicker}, S.~R. and {D{\"u}nner}, R. and {Fowler}, J.~W. and
         {Fox}, A.~E. and {Gallardo}, P.~A. and {Gao}, J. and {Grace}, E. and
         {Halpern}, M. and {Hasselfield}, M. and {Henderson}, S.~W. and
         {Hilton}, G.~C. and {Hincks}, A.~D. and {Ho}, S.~P. and {Hubmayr}, J. and
         {Irwin}, K.~D. and {Klein}, J. and {Koopman}, B. and {Li}, Dale and
         {Louis}, T. and {Lungu}, M. and {Maurin}, L. and {McMahon}, J. and
         {Munson}, C.~D. and {Naess}, S. and {Nati}, F. and {Newburgh}, L. and
         {Nibarger}, J. and {Niemack}, M.~D. and {Niraula}, P. and
         {Nolta}, M.~R. and {Page}, L.~A. and {Pappas}, C.~G. and
         {Schillaci}, A. and {Schmitt}, B.~L. and {Sehgal}, N. and
         {Sievers}, J.~L. and {Simon}, S.~M. and {Staggs}, S.~T. and
         {Tucker}, C. and {Uehara}, M. and {van Lanen}, J. and {Ward}, J.~T. and
         {Wollack}, E.~J.},
        title = "{The Atacama Cosmology Telescope: The Polarization-sensitive ACTPol Instrument}",
      journal = {\apjs},
         year = 2016,
        month = dec,
       volume = {227},
       number = {2},
          eid = {21},
        pages = {21},
          doi = {10.3847/1538-4365/227/2/21},
archivePrefix = {arXiv},
       eprint = {1605.06569},
 primaryClass = {astro-ph.IM},
       adsurl = {https://ui.adsabs.harvard.edu/abs/2016ApJS..227...21T}
}

@ARTICLE{Coble2007,
       author = {{Coble}, K. and {Bonamente}, M. and {Carlstrom}, J.~E. and {Dawson}, K. and {Hasler}, N. and {Holzapfel}, W. and {Joy}, M. and {La Roque}, S. and {Marrone}, D.~P. and {Reese}, E.~D.},
        title = "{Radio Sources toward Galaxy Clusters at 30 GHz}",
      journal = {\aj},
         year = 2007,
        month = sep,
       volume = {134},
       number = {3},
        pages = {897},
          doi = {10.1086/519973},
archivePrefix = {arXiv},
       eprint = {astro-ph/0608274},
 primaryClass = {astro-ph},
       adsurl = {https://ui.adsabs.harvard.edu/abs/2007AJ....134..897C}
}

@article{Dicker_2024,
   title={Sensitive 3 mm Imaging of Discrete Sources in the Fields of Thermal Sunyaev–Zel’dovich Effect–Selected Galaxy Clusters},
   volume={970},
   ISSN={1538-4357},
   url={http://dx.doi.org/10.3847/1538-4357/ad4e35},
   DOI={10.3847/1538-4357/ad4e35},
   number={1},
   journal={The Astrophysical Journal},
   publisher={American Astronomical Society},
   author={Dicker, Simon R. and Sarmiento, Karen Perez and Mason, Brian and Bhandarkar, Tanay and Devlin, Mark J. and Di Mascolo, Luca and Haridas, Saianeesh and Hilton, Matt and Madhavacheril, Mathew and Moravec, Emily and Mroczkowski, Tony and Orlowski-Scherer, John and Romero, Charles and Sarazin, Craig L. and Sievers, Jonathan},
   year={2024},
   month=jul, pages={84} }

@ARTICLE{Gralla2020,
       author = {{Gralla}, Megan B. and {Marriage}, Tobias A.},
        title = "{Accounting for Selection Bias Using Simulations: A General Method and an Application to Millimeter-wavelength Surveys}",
      journal = {\apj},
         year = 2020,
        month = apr,
       volume = {893},
       number = {2},
          eid = {103},
        pages = {103},
          doi = {10.3847/1538-4357/ab7916},
archivePrefix = {arXiv},
       eprint = {1905.04593},
 primaryClass = {astro-ph.GA},
       adsurl = {https://ui.adsabs.harvard.edu/abs/2020ApJ...893..103G}
}

@ARTICLE{MroczkowskiSZReview,
       author = {{Mroczkowski}, Tony and {Nagai}, Daisuke and {Basu}, Kaustuv and
         {Chluba}, Jens and {Sayers}, Jack and {Adam}, R{\'e}mi and
         {Churazov}, Eugene and {Crites}, Abigail and {Di Mascolo}, Luca and
         {Eckert}, Dominique and {Macias-Perez}, Juan and
         {Mayet}, Fr{\'e}d{\'e}ric and {Perotto}, Laurence and
         {Pointecouteau}, Etienne and {Romero}, Charles and {Ruppin}, Florian and
         {Scannapieco}, Evan and {ZuHone}, John},
        title = "{Astrophysics with the Spatially and Spectrally Resolved Sunyaev-Zeldovich Effects. A Millimetre/Submillimetre Probe of the Warm and Hot Universe}",
      journal = {\ssr},
         year = "2019",
        month = "Feb",
       volume = {215},
       number = {1},
          eid = {17},
        pages = {17},
          doi = {10.1007/s11214-019-0581-2},
archivePrefix = {arXiv},
       eprint = {1811.02310},
 primaryClass = {astro-ph.CO},
       adsurl = {https://ui.adsabs.harvard.edu/abs/2019SSRv..215...17M}
}

@ARTICLE{Romero2020,
       author = {{Romero}, Charles E. and {Sievers}, Jonathan and {Ghirardini}, Vittorio and
         {Dicker}, Simon and {Giacintucci}, Simona and {\bf Mroczkowski}, Tony and
         {Mason}, Brian S. and {Sarazin}, Craig and {Devlin}, Mark and
         {Gaspari}, Massimo and {Battaglia}, Nicholas and {Hilton}, Matthew and
         {Bulbul}, Esra and {Lowe}, Ian and {Stanchfield}, Sara},
        title = "{Pressure Profiles and Mass Estimates Using High-resolution Sunyaev-Zel'dovich Effect Observations of Zwicky 3146 with MUSTANG-2}",
      journal = {\apj},
         year = 2020,
        month = mar,
       volume = {891},
       number = {1},
          eid = {90},
        pages = {90},
          doi = {10.3847/1538-4357/ab6d70},
archivePrefix = {arXiv},
       eprint = {1908.09200},
 primaryClass = {astro-ph.CO},
       adsurl = {https://ui.adsabs.harvard.edu/abs/2020ApJ...891...90R}
}

@INPROCEEDINGS{SPIRE,
       author = {{Schulz}, Bernhard and {Marton}, G{\'a}bor and {Valtchanov}, Ivan and {P{\'e}rez Garc{\'\i}a}, Ana Mar{\'\i}a and {Pint{\'e}r}, S{\'a}ndor and {Appleton}, Phil and {Kiss}, Csaba and {Lim}, Tanya and {Lu}, Nanyao and {Papageorgiou}, Andreas and {Pearson}, Chris and {Rector}, John and {S{\'a}nchez Portal}, Miguel and {Shupe}, David and {T{\'o}th}, Viktor L. and {Van Dyk}, Schuyler and {Varga-Vereb{\'e}lyi}, Erika and {Xu}, Kevin},
        title = "{The Herschel-SPIRE Point Source Catalog Version 2}",
    booktitle = {American Astronomical Society Meeting Abstracts \#231},
         year = 2018,
       series = {American Astronomical Society Meeting Abstracts},
       volume = {231},
        month = jan,
          eid = {361.21},
        pages = {361.21},
       adsurl = {https://ui.adsabs.harvard.edu/abs/2018AAS...23136121S}
}

@ARTICLE{TGSS,
       author = {{Intema}, H.~T. and {Jagannathan}, P. and {Mooley}, K.~P. and {Frail}, D.~A.},
        title = "{The GMRT 150 MHz all-sky radio survey. First alternative data release TGSS ADR1}",
      journal = {\aap},
         year = 2017,
        month = feb,
       volume = {598},
          eid = {A78},
        pages = {A78},
          doi = {10.1051/0004-6361/201628536},
archivePrefix = {arXiv},
       eprint = {1603.04368},
 primaryClass = {astro-ph.CO},
       adsurl = {https://ui.adsabs.harvard.edu/abs/2017A&A...598A..78I}
}

@ARTICLE{FIRST,
       author = {{White}, Richard L. and {Becker}, Robert H. and {Helfand}, David J. and {Gregg}, Michael D.},
        title = "{A Catalog of 1.4 GHz Radio Sources from the FIRST Survey}",
      journal = {\apj},
         year = 1997,
        month = feb,
       volume = {475},
       number = {2},
        pages = {479-493},
          doi = {10.1086/303564},
       adsurl = {https://ui.adsabs.harvard.edu/abs/1997ApJ...475..479W}
}

@ARTICLE{VLASS,
       author = {{Gordon}, Yjan A. and {Boyce}, Michelle M. and {O'Dea}, Christopher P. and {Rudnick}, Lawrence and {Andernach}, Heinz and {Vantyghem}, Adrian N. and {Baum}, Stefi A. and {Bui}, Jean-Paul and {Dionyssiou}, Mathew and {Safi-Harb}, Samar and {Sander}, Isabel},
        title = "{A Quick Look at the 3 GHz Radio Sky. I. Source Statistics from the Very Large Array Sky Survey}",
      journal = {\apjs},
         year = 2021,
        month = aug,
       volume = {255},
       number = {2},
          eid = {30},
        pages = {30},
          doi = {10.3847/1538-4365/ac05c0},
archivePrefix = {arXiv},
       eprint = {2102.11753},
 primaryClass = {astro-ph.GA},
       adsurl = {https://ui.adsabs.harvard.edu/abs/2021ApJS..255...30G}
}

@ARTICLE{Lui2022,
       author = {{Liu}, A. and {Bulbul}, E. and {Ghirardini}, V. and {Liu}, T. and {Klein}, M. and {Clerc}, N. and {{\"O}zsoy}, Y. and {Ramos-Ceja}, M.~E. and {Pacaud}, F. and {Comparat}, J. and {Okabe}, N. and {Bahar}, Y.~E. and {Biffi}, V. and {Brunner}, H. and {Br{\"u}ggen}, M. and {Buchner}, J. and {Ider Chitham}, J. and {Chiu}, I. and {Dolag}, K. and {Gatuzz}, E. and {Gonzalez}, J. and {Hoang}, D.~N. and {Lamer}, G. and {Merloni}, A. and {Nandra}, K. and {Oguri}, M. and {Ota}, N. and {Predehl}, P. and {Reiprich}, T.~H. and {Salvato}, M. and {Schrabback}, T. and {Sanders}, J.~S. and {Seppi}, R. and {Thibaud}, Q.},
        title = "{The eROSITA Final Equatorial-Depth Survey (eFEDS). Catalog of galaxy clusters and groups}",
      journal = {\aap},
         year = 2022,
        month = may,
       volume = {661},
          eid = {A2},
        pages = {A2},
          doi = {10.1051/0004-6361/202141120},
archivePrefix = {arXiv},
       eprint = {2106.14518},
 primaryClass = {astro-ph.CO},
       adsurl = {https://ui.adsabs.harvard.edu/abs/2022A&A...661A...2L}
}

@ARTICLE{Predehl_2021,
       author = {{Predehl}, P. and {Andritschke}, R. and {Arefiev}, V. and {Babyshkin}, V. and {Batanov}, O. and {Becker}, W. and {B{\"o}hringer}, H. and {Bogomolov}, A. and {Boller}, T. and {Borm}, K. and {Bornemann}, W. and {Br{\"a}uninger}, H. and {Br{\"u}ggen}, M. and {Brunner}, H. and {Brusa}, M. and {Bulbul}, E. and {Buntov}, M. and {Burwitz}, V. and {Burkert}, W. and {Clerc}, N. and {Churazov}, E. and {Coutinho}, D. and {Dauser}, T. and {Dennerl}, K. and {Doroshenko}, V. and {Eder}, J. and {Emberger}, V. and {Eraerds}, T. and {Finoguenov}, A. and {Freyberg}, M. and {Friedrich}, P. and {Friedrich}, S. and {F{\"u}rmetz}, M. and {Georgakakis}, A. and {Gilfanov}, M. and {Granato}, S. and {Grossberger}, C. and {Gueguen}, A. and {Gureev}, P. and {Haberl}, F. and {H{\"a}lker}, O. and {Hartner}, G. and {Hasinger}, G. and {Huber}, H. and {Ji}, L. and {Kienlin}, A. v. and {Kink}, W. and {Korotkov}, F. and {Kreykenbohm}, I. and {Lamer}, G. and {Lomakin}, I. and {Lapshov}, I. and {Liu}, T. and {Maitra}, C. and {Meidinger}, N. and {Menz}, B. and {Merloni}, A. and {Mernik}, T. and {Mican}, B. and {Mohr}, J. and {M{\"u}ller}, S. and {Nandra}, K. and {Nazarov}, V. and {Pacaud}, F. and {Pavlinsky}, M. and {Perinati}, E. and {Pfeffermann}, E. and {Pietschner}, D. and {Ramos-Ceja}, M.~E. and {Rau}, A. and {Reiffers}, J. and {Reiprich}, T.~H. and {Robrade}, J. and {Salvato}, M. and {Sanders}, J. and {Santangelo}, A. and {Sasaki}, M. and {Scheuerle}, H. and {Schmid}, C. and {Schmitt}, J. and {Schwope}, A. and {Shirshakov}, A. and {Steinmetz}, M. and {Stewart}, I. and {Str{\"u}der}, L. and {Sunyaev}, R. and {Tenzer}, C. and {Tiedemann}, L. and {Tr{\"u}mper}, J. and {Voron}, V. and {Weber}, P. and {Wilms}, J. and {Yaroshenko}, V.},
        title = "{The eROSITA X-ray telescope on SRG}",
      journal = {\aap},
         year = 2021,
        month = mar,
       volume = {647},
          eid = {A1},
        pages = {A1},
          doi = {10.1051/0004-6361/202039313},
archivePrefix = {arXiv},
       eprint = {2010.03477},
 primaryClass = {astro-ph.HE},
       adsurl = {https://ui.adsabs.harvard.edu/abs/2021A&A...647A...1P}
}

@ARTICLE{Dicker2021,
    author = {Dicker, Simon R and Battistelli, Elia S and Bhandarkar, Tanay and Devlin, Mark J and Duff, Shannon M and Hilton, Gene and Hilton, Matt and Hincks, Adam D and Hubmayr, Johannes and Huffenberger, Kevin and Hughes, John P and Di Mascolo, Luca and Mason, Brian S and Mates, J A B and McMahon, Jeff and Mroczkowski, Tony and Naess, Sigurd and Orlowski-Scherer, John and Partridge, Bruce and Radiconi, Federico and Romero, Charles and Sarazin, Craig L and Sehgal, Neelima and Sievers, Jonathan and Sifón, Cristóbal and Ullom, Joel and Vale, Leila R and Vissers, Michael R and Xu, Zhilei},
    title = "{Observations of compact sources in galaxy clusters using MUSTANG2}",
    journal = {Monthly Notices of the Royal Astronomical Society},
    volume = {508},
    number = {2},
    pages = {2600-2612},
    year = {2021},
    month = {09},
    issn = {0035-8711},
    doi = {10.1093/mnras/stab2679},
    url = {https://doi.org/10.1093/mnras/stab2679},
    eprint = {https://academic.oup.com/mnras/article-pdf/508/2/2600/40587593/stab2679.pdf},
}

@ARTICLE{Li2022,
       author = {{Li}, Zack and {Puglisi}, Giuseppe and {Madhavacheril}, Mathew S. and {Alvarez}, Marcelo A.},
        title = "{Simulated catalogs and maps of radio galaxies at millimeter wavelengths in Websky}",
      journal = {\jcap},
         year = 2022,
        month = aug,
       volume = {2022},
       number = {8},
          eid = {029},
        pages = {029},
          doi = {10.1088/1475-7516/2022/08/029},
archivePrefix = {arXiv},
       eprint = {2110.15357},
 primaryClass = {astro-ph.GA},
       adsurl = {https://ui.adsabs.harvard.edu/abs/2022JCAP...08..029L}
}

@ARTICLE{SimonsForecastPaper,
   title={The Simons Observatory: science goals and forecasts},
   volume={2019},
   ISSN={1475-7516},
   url={http://dx.doi.org/10.1088/1475-7516/2019/02/056},
   DOI={10.1088/1475-7516/2019/02/056},
   number={02},
   journal={Journal of Cosmology and Astroparticle Physics},
   publisher={IOP Publishing},
   author={Ade, Peter and Aguirre, James and Ahmed, Zeeshan and Aiola, Simone and Ali, Aamir and Alonso, David and Alvarez, Marcelo A. and Arnold, Kam and Ashton, Peter and Austermann, Jason and et al.},
   year={2019},
   month={Feb},
   pages={056–056}
}

@ARTICLE{Hilton2020,
      author = {{Hilton}, M. and {Sif{\'o}n}, C. and {Naess}, S. and {Madhavacheril}, M. and {Oguri}, M. and {Rozo}, E. and {Rykoff}, E. and {Abbott}, T.~M.~C. and {Adhikari}, S. and {Aguena}, M. and {Aiola}, S. and {Allam}, S. and {Amodeo}, S. and {Amon}, A. and {Annis}, J. and {Ansarinejad}, B. and {Aros-Bunster}, C. and {Austermann}, J.~E. and {Avila}, S. and {Bacon}, D. and {Battaglia}, N. and {Beall}, J.~A. and {Becker}, D.~T. and {Bernstein}, G.~M. and {Bertin}, E. and {Bhandarkar}, T. and {Bhargava}, S. and {Bond}, J.~R. and {Brooks}, D. and {Burke}, D.~L. and {Calabrese}, E. and {Carrasco Kind}, M. and {Carretero}, J. and {Choi}, S.~K. and {Choi}, A. and {Conselice}, C. and {da Costa}, L.~N. and {Costanzi}, M. and {Crichton}, D. and {Crowley}, K.~T. and {D{\"u}nner}, R. and {Denison}, E.~V. and {Devlin}, M.~J. and {Dicker}, S.~R. and {Diehl}, H.~T. and {Dietrich}, J.~P. and {Doel}, P. and {Duff}, S.~M. and {Duivenvoorden}, A.~J. and {Dunkley}, J. and {Everett}, S. and {Ferraro}, S. and {Ferrero}, I. and {Fert{\'e}}, A. and {Flaugher}, B. and {Frieman}, J. and {Gallardo}, P.~A. and {Garc{\'\i}a-Bellido}, J. and {Gaztanaga}, E. and {Gerdes}, D.~W. and {Giles}, P. and {Golec}, J.~E. and {Gralla}, M.~B. and {Grandis}, S. and {Gruen}, D. and {Gruendl}, R.~A. and {Gschwend}, J. and {Gutierrez}, G. and {Han}, D. and {Hartley}, W.~G. and {Hasselfield}, M. and {Hill}, J.~C. and {Hilton}, G.~C. and {Hincks}, A.~D. and {Hinton}, S.~R. and {Ho}, S. -P.~P. and {Honscheid}, K. and {Hoyle}, B. and {Hubmayr}, J. and {Huffenberger}, K.~M. and {Hughes}, J.~P. and {Jaelani}, A.~T. and {Jain}, B. and {James}, D.~J. and {Jeltema}, T. and {Kent}, S. and {Knowles}, K. and {Koopman}, B.~J. and {Kuehn}, K. and {Lahav}, O. and {Lima}, M. and {Lin}, Y. -T. and {Lokken}, M. and {Loubser}, S.~I. and {MacCrann}, N. and {Maia}, M.~A.~G. and {Marriage}, T.~A. and {Martin}, J. and {McMahon}, J. and {Melchior}, P. and {Menanteau}, F. and {Miquel}, R. and {Miyatake}, H. and {Moodley}, K. and {Morgan}, R. and {Mroczkowski}, T. and {Nati}, F. and {Newburgh}, L.~B. and {Niemack}, M.~D. and {Nishizawa}, A.~J. and {Ogando}, R.~L.~C. and {Orlowski-Scherer}, J. and {Page}, L.~A. and {Palmese}, A. and {Partridge}, B. and {Paz-Chinch{\'o}n}, F. and {Phakathi}, P. and {Plazas}, A.~A. and {Robertson}, N.~C. and {Romer}, A.~K. and {Carnero Rosell}, A. and {Salatino}, M. and {Sanchez}, E. and {Schaan}, E. and {Schillaci}, A. and {Sehgal}, N. and {Serrano}, S. and {Shin}, T. and {Simon}, S.~M. and {Smith}, M. and {Soares-Santos}, M. and {Spergel}, D.~N. and {Staggs}, S.~T. and {Storer}, E.~R. and {Suchyta}, E. and {Swanson}, M.~E.~C. and {Tarle}, G. and {Thomas}, D. and {To}, C. and {Trac}, H. and {Ullom}, J.~N. and {Vale}, L.~R. and {Van Lanen}, J. and {Vavagiakis}, E.~M. and {De Vicente}, J. and {Wilkinson}, R.~D. and {Wollack}, E.~J. and {Xu}, Z. and {Zhang}, Y.},
        title = "{The Atacama Cosmology Telescope: A Catalog of $>$4000 Sunyaev-Zel{\textquoteright}dovich Galaxy Clusters}",
      journal = {\apjs},
         year = 2021,
        month = mar,
       volume = {253},
       number = {1},
          eid = {3},
        pages = {3},
          doi = {10.3847/1538-4365/abd023},
archivePrefix = {arXiv},
       eprint = {2009.11043},
 primaryClass = {astro-ph.CO},
       adsurl = {https://ui.adsabs.harvard.edu/abs/2021ApJS..253....3H}
}

@ARTICLE{wl_eFEDS,
       author = {{Chiu}, I.-Non and {Ghirardini}, Vittorio and {Liu}, Ang and {Grandis}, Sebastian and {Bulbul}, Esra and {Bahar}, Y. Emre and {Comparat}, Johan and {Bocquet}, Sebastian and {Clerc}, Nicolas and {Klein}, Matthias and {Liu}, Teng and {Li}, Xiangchong and {Miyatake}, Hironao and {Mohr}, Joseph and {More}, Surhud and {Oguri}, Masamune and {Okabe}, Nobuhiro and {Pacaud}, Florian and {Ramos-Ceja}, Miriam E. and {Reiprich}, Thomas H. and {Schrabback}, Tim and {Umetsu}, Keiichi},
        title = "{The eROSITA Final Equatorial-Depth Survey (eFEDS). X-ray observable-to-mass-and-redshift relations of galaxy clusters and groups with weak-lensing mass calibration from the Hyper Suprime-Cam Subaru Strategic Program survey}",
      journal = {\aap},
         year = 2022,
        month = may,
       volume = {661},
          eid = {A11},
        pages = {A11},
          doi = {10.1051/0004-6361/202141755},
archivePrefix = {arXiv},
       eprint = {2107.05652},
 primaryClass = {astro-ph.CO},
       adsurl = {https://ui.adsabs.harvard.edu/abs/2022A&A...661A..11C}
}

@ARTICLE{Stein2020,
       author = {{Stein}, George and {Alvarez}, Marcelo A. and {Bond}, J. Richard and {van Engelen}, Alexander and {Battaglia}, Nicholas},
        title = "{The Websky extragalactic CMB simulations}",
      journal = {\jcap},
         year = 2020,
        month = oct,
       volume = {2020},
       number = {10},
          eid = {012},
        pages = {012},
          doi = {10.1088/1475-7516/2020/10/012},
archivePrefix = {arXiv},
       eprint = {2001.08787},
 primaryClass = {astro-ph.CO},
       adsurl = {https://ui.adsabs.harvard.edu/abs/2020JCAP...10..012S}
}

@ARTICLE{2018AJ....156..123A,
       author = {{Astropy Collaboration} and {Price-Whelan}, A.~M. and {Sip{\H{o}}cz}, B.~M. and {G{\"u}nther}, H.~M. and {Lim}, P.~L. and {Crawford}, S.~M. and {Conseil}, S. and {Shupe}, D.~L. and {Craig}, M.~W. and {Dencheva}, N. and {Ginsburg}, A. and {VanderPlas}, J.~T. and {Bradley}, L.~D. and {P{\'e}rez-Su{\'a}rez}, D. and {de Val-Borro}, M. and {Aldcroft}, T.~L. and {Cruz}, K.~L. and {Robitaille}, T.~P. and {Tollerud}, E.~J. and {Ardelean}, C. and {Babej}, T. and {Bach}, Y.~P. and {Bachetti}, M. and {Bakanov}, A.~V. and {Bamford}, S.~P. and {Barentsen}, G. and {Barmby}, P. and {Baumbach}, A. and {Berry}, K.~L. and {Biscani}, F. and {Boquien}, M. and {Bostroem}, K.~A. and {Bouma}, L.~G. and {Brammer}, G.~B. and {Bray}, E.~M. and {Breytenbach}, H. and {Buddelmeijer}, H. and {Burke}, D.~J. and {Calderone}, G. and {Cano Rodr{\'\i}guez}, J.~L. and {Cara}, M. and {Cardoso}, J.~V.~M. and {Cheedella}, S. and {Copin}, Y. and {Corrales}, L. and {Crichton}, D. and {D'Avella}, D. and {Deil}, C. and {Depagne}, {\'E}. and {Dietrich}, J.~P. and {Donath}, A. and {Droettboom}, M. and {Earl}, N. and {Erben}, T. and {Fabbro}, S. and {Ferreira}, L.~A. and {Finethy}, T. and {Fox}, R.~T. and {Garrison}, L.~H. and {Gibbons}, S.~L.~J. and {Goldstein}, D.~A. and {Gommers}, R. and {Greco}, J.~P. and {Greenfield}, P. and {Groener}, A.~M. and {Grollier}, F. and {Hagen}, A. and {Hirst}, P. and {Homeier}, D. and {Horton}, A.~J. and {Hosseinzadeh}, G. and {Hu}, L. and {Hunkeler}, J.~S. and {Ivezi{\'c}}, {\v{Z}}. and {Jain}, A. and {Jenness}, T. and {Kanarek}, G. and {Kendrew}, S. and {Kern}, N.~S. and {Kerzendorf}, W.~E. and {Khvalko}, A. and {King}, J. and {Kirkby}, D. and {Kulkarni}, A.~M. and {Kumar}, A. and {Lee}, A. and {Lenz}, D. and {Littlefair}, S.~P. and {Ma}, Z. and {Macleod}, D.~M. and {Mastropietro}, M. and {McCully}, C. and {Montagnac}, S. and {Morris}, B.~M. and {Mueller}, M. and {Mumford}, S.~J. and {Muna}, D. and {Murphy}, N.~A. and {Nelson}, S. and {Nguyen}, G.~H. and {Ninan}, J.~P. and {N{\"o}the}, M. and {Ogaz}, S. and {Oh}, S. and {Parejko}, J.~K. and {Parley}, N. and {Pascual}, S. and {Patil}, R. and {Patil}, A.~A. and {Plunkett}, A.~L. and {Prochaska}, J.~X. and {Rastogi}, T. and {Reddy Janga}, V. and {Sabater}, J. and {Sakurikar}, P. and {Seifert}, M. and {Sherbert}, L.~E. and {Sherwood-Taylor}, H. and {Shih}, A.~Y. and {Sick}, J. and {Silbiger}, M.~T. and {Singanamalla}, S. and {Singer}, L.~P. and {Sladen}, P.~H. and {Sooley}, K.~A. and {Sornarajah}, S. and {Streicher}, O. and {Teuben}, P. and {Thomas}, S.~W. and {Tremblay}, G.~R. and {Turner}, J.~E.~H. and {Terr{\'o}n}, V. and {van Kerkwijk}, M.~H. and {de la Vega}, A. and {Watkins}, L.~L. and {Weaver}, B.~A. and {Whitmore}, J.~B. and {Woillez}, J. and {Zabalza}, V. and {Astropy Contributors}},
        title = "{The Astropy Project: Building an Open-science Project and Status of the v2.0 Core Package}",
      journal = {\aj},
         year = 2018,
        month = sep,
       volume = {156},
       number = {3},
          eid = {123},
        pages = {123},
          doi = {10.3847/1538-3881/aabc4f},
archivePrefix = {arXiv},
       eprint = {1801.02634},
 primaryClass = {astro-ph.IM},
       adsurl = {https://ui.adsabs.harvard.edu/abs/2018AJ....156..123A}
}

@ARTICLE{2013A&A...558A..33A,
       author = {{Astropy Collaboration} and {Robitaille}, Thomas P. and
         {Tollerud}, Erik J. and {Greenfield}, Perry and {Droettboom}, Michael and
         {Bray}, Erik and {Aldcroft}, Tom and {Davis}, Matt and
         {Ginsburg}, Adam and {Price-Whelan}, Adrian M. and
         {Kerzendorf}, Wolfgang E. and {Conley}, Alexander and {Crighton}, Neil and
         {Barbary}, Kyle and {Muna}, Demitri and {Ferguson}, Henry and
         {Grollier}, Fr{\'e}d{\'e}ric and {Parikh}, Madhura M. and
         {Nair}, Prasanth H. and {Unther}, Hans M. and {Deil}, Christoph and
         {Woillez}, Julien and {Conseil}, Simon and {Kramer}, Roban and
         {Turner}, James E.~H. and {Singer}, Leo and {Fox}, Ryan and
         {Weaver}, Benjamin A. and {Zabalza}, Victor and {Edwards}, Zachary I. and
         {Azalee Bostroem}, K. and {Burke}, D.~J. and {Casey}, Andrew R. and
         {Crawford}, Steven M. and {Dencheva}, Nadia and {Ely}, Justin and
         {Jenness}, Tim and {Labrie}, Kathleen and {Lim}, Pey Lian and
         {Pierfederici}, Francesco and {Pontzen}, Andrew and {Ptak}, Andy and
         {Refsdal}, Brian and {Servillat}, Mathieu and {Streicher}, Ole},
        title = "{Astropy: A community Python package for astronomy}",
      journal = {\aap},
         year = "2013",
        month = "Oct",
       volume = {558},
          eid = {A33},
        pages = {A33},
          doi = {10.1051/0004-6361/201322068},
archivePrefix = {arXiv},
       eprint = {1307.6212},
 primaryClass = {astro-ph.IM},
       adsurl = {https://ui.adsabs.harvard.edu/abs/2013A&A...558A..33A}
}

@MISC{vo:WISE_scs,
  year=2013,
  title={{WISE} All-Sky Release Catalog {SCS}},
  author={Wright, E.L. and Cutri, R.C. and et al},
  url={http://dc.zah.uni-heidelberg.de/wise/q/s/info},
  howpublished={{VO} resource provided by the {GAVO} Data Center}
}

@MISC{WISE_pntsrc,
       author = {{Cutri}, R.~M. and {Wright}, E.~L. and {Conrow}, T. and {Bauer}, J. and {Benford}, D. and {Brandenburg}, H. and {Dailey}, J. and {Eisenhardt}, P.~R.~M. and {Evans}, T. and {Fajardo-Acosta}, S. and {Fowler}, J. and {Gelino}, C. and {Grillmair}, C. and {Harbut}, M. and {Hoffman}, D. and {Jarrett}, T. and {Kirkpatrick}, J.~D. and {Leisawitz}, D. and {Liu}, W. and {Mainzer}, A. and {Marsh}, K. and {Masci}, F. and {McCallon}, H. and {Padgett}, D. and {Ressler}, M.~E. and {Royer}, D. and {Skrutskie}, M.~F. and {Stanford}, S.~A. and {Wyatt}, P.~L. and {Tholen}, D. and {Tsai}, C.~W. and {Wachter}, S. and {Wheelock}, S.~L. and {Yan}, L. and {Alles}, R. and {Beck}, R. and {Grav}, T. and {Masiero}, J. and {McCollum}, B. and {McGehee}, P. and {Papin}, M. and {Wittman}, M.},
        title = "{Explanatory Supplement to the WISE All-Sky Data Release Products}",
    year = 2012,
        month = mar,
        pages = {1},
       adsurl = {https://ui.adsabs.harvard.edu/abs/2012wise.rept....1C}
}
\bibliographystyle{aasjournal}

%astropH
\appendix
%uncomment this line to produce an expanded version of Fig 4 
%\setcounter{page}{1}

\renewcommand{\thefigure}{4a}
  \begin{figure*}[tb]
    \centering
    \includegraphics[width=0.46\linewidth]{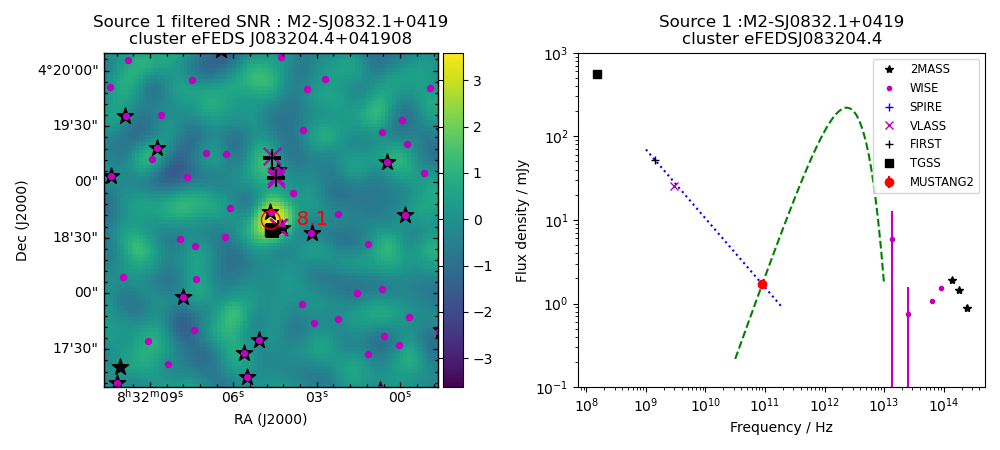}\hspace{0.8cm}
    \includegraphics[width=0.46\linewidth]{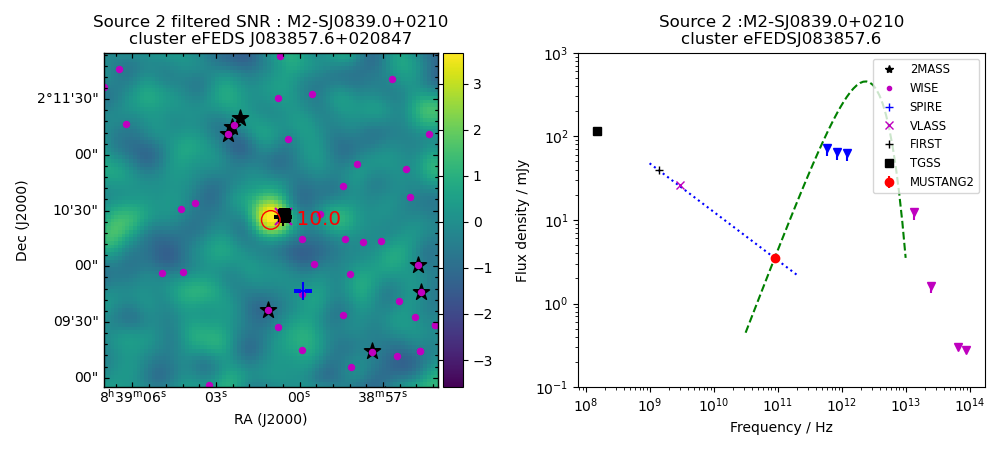}
    
   \includegraphics[width=0.46\linewidth]{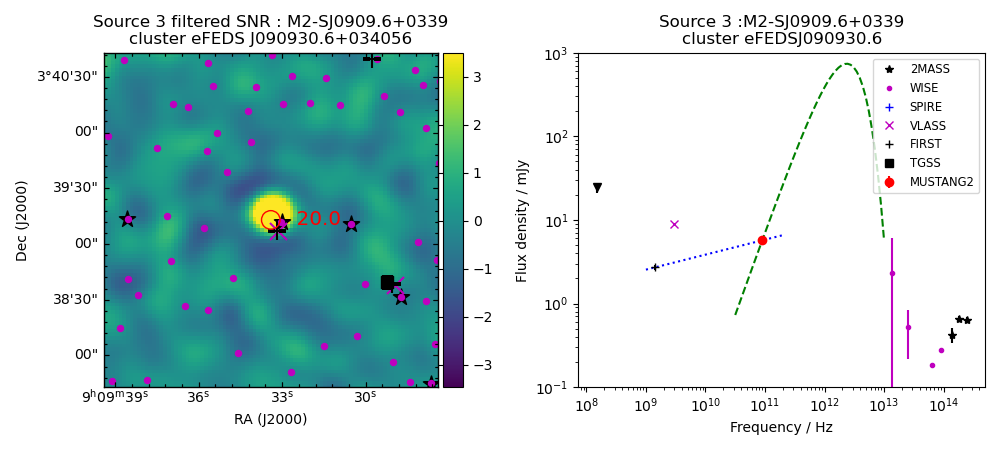}\hspace{0.8cm}
    \includegraphics[width=0.46\linewidth]{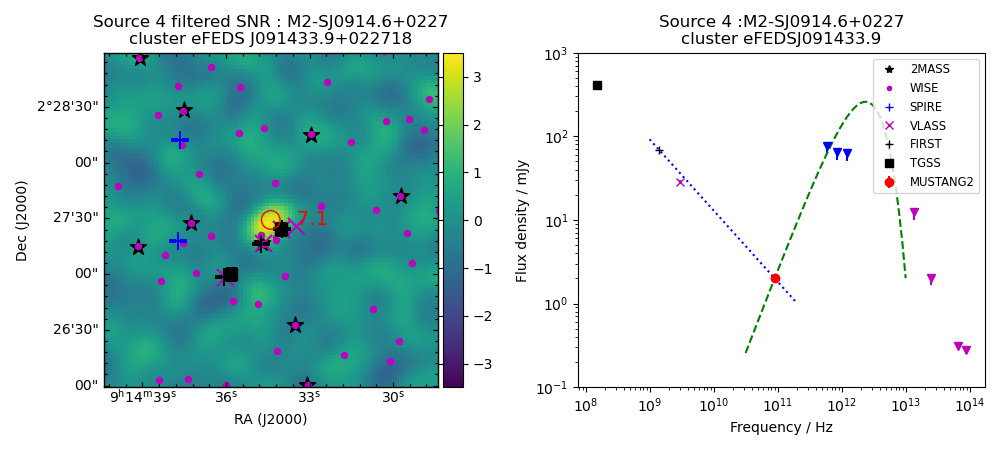}
    
    \includegraphics[width=0.46\linewidth]{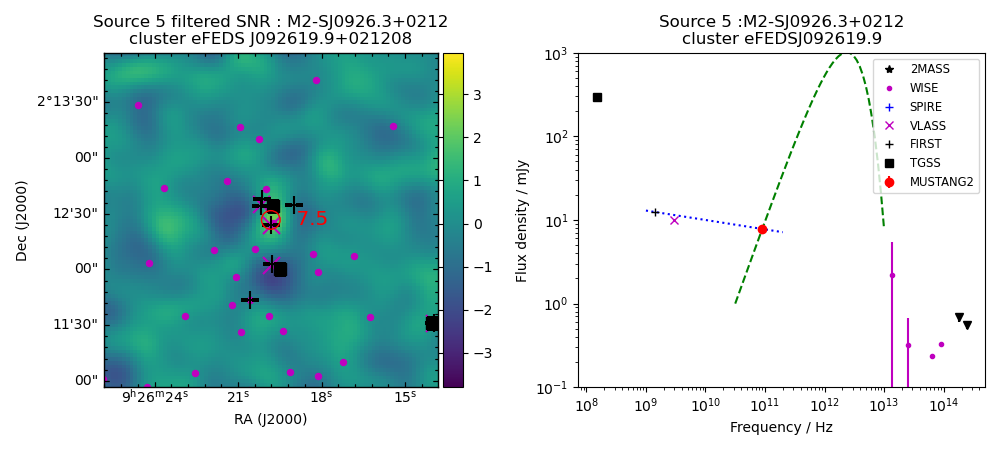}\hspace{0.8cm}
    \includegraphics[width=0.46\linewidth]{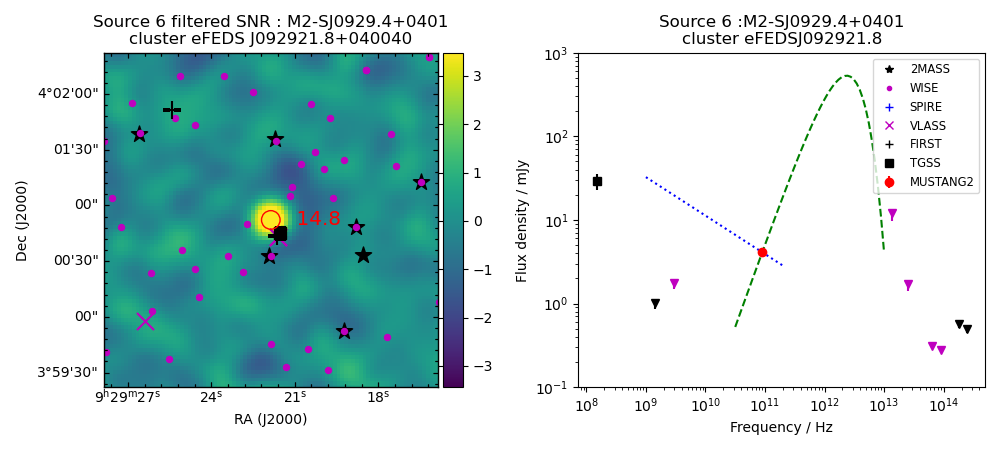}\vspace{0.08cm}

    \caption{On-line Extra: \figfourcaption \newline
    %notes on individual clusters can go here
    }
\end{figure*}

\renewcommand{\thefigure}{4b}
\begin{figure*}[tb]
    \centering
     \includegraphics[width=0.46\linewidth]{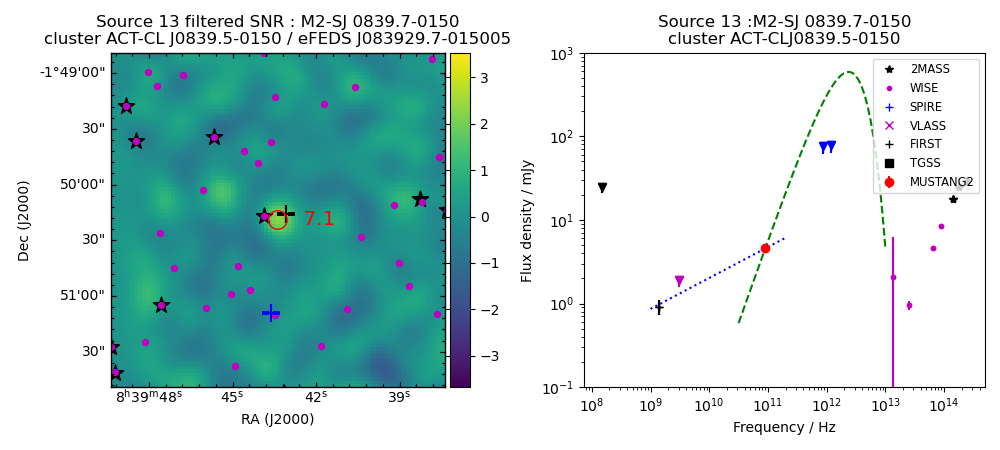}\hspace{0.8cm}
    \includegraphics[width=0.46\linewidth]{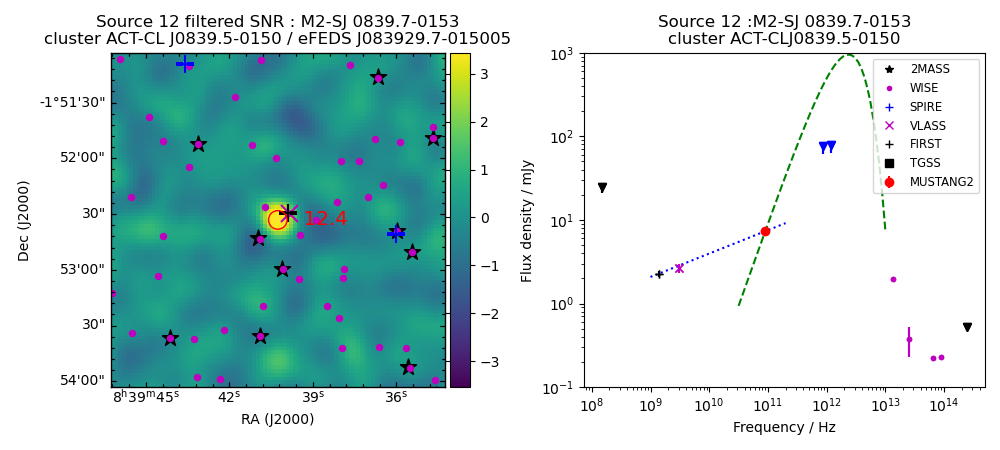}
    
   \includegraphics[width=0.46\linewidth]{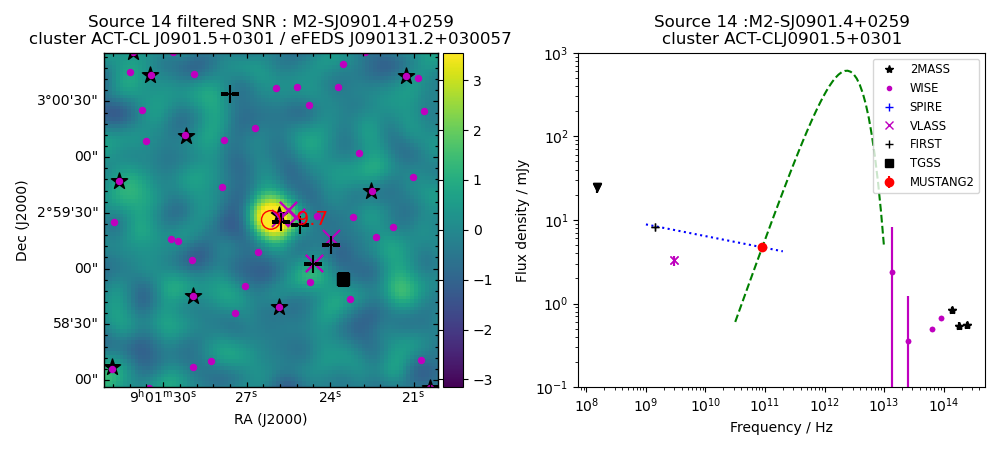}\hspace{0.8cm}
    \includegraphics[width=0.46\linewidth]{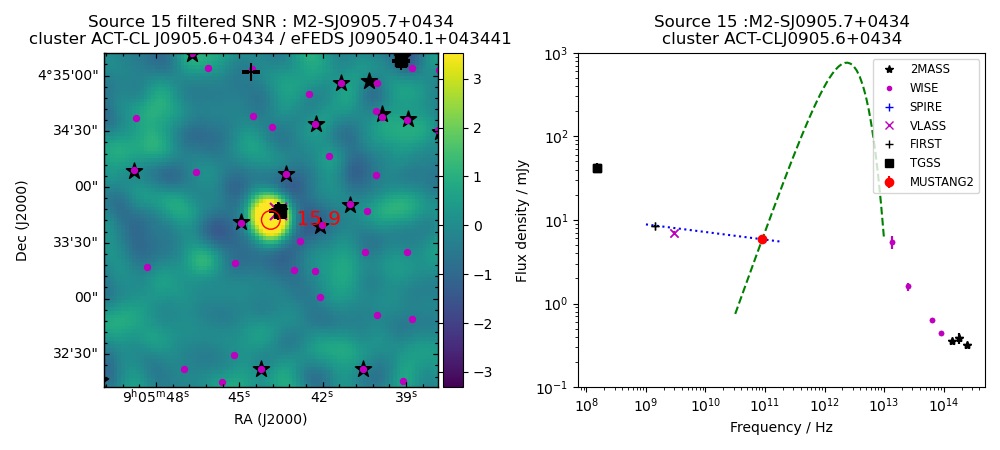}
    
   \includegraphics[width=0.46\linewidth]{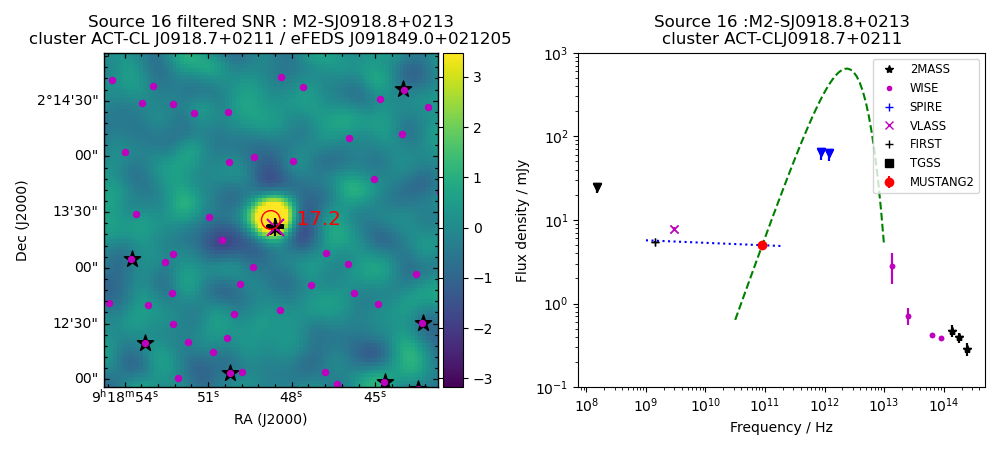}%\hspace{0.8cm}
    %\includegraphics[width=0.46\linewidth]{snapshots_DR5/fig4set-M2-SJ09}
      
    %\captionsetup{labelformat=empty}
    %\includegraphics[width=\linewidth]{figures/SED_plots/ACT-CLJ0201.6-0211_ra_30.429771368113496_dec_-2.19770479160654.png}
    \caption{On-line Extra: Sources found in \cite{Dicker_2024} that are also eFEDS clusters. \figfourcaption \newline}
    %notes on individual clusters can go here}
\end{figure*}

%Uncomment this line to include a full version of table 1, will be submitted to the journal as a text file.
\setlength{\LTcapwidth}{0.8\textwidth}
\begin{longtable}{lccccc}
    %\centering
    \caption{A full list of the clusters included in the X-ray sample along with the noise in the center of each map, their masses estimated by different methods, and their redshifts.  Clusters marked with a $^*$ were also detected in the tSZ and included in the D2024 sample. For other clusters, the tSZ masses were found using forced photometry at the location of the eFEDS cluster and make use of the same weak lensing scaling relationship used in DR5. In many cases this resulted in a negative compton Y. For those clusters with measured sources, two tSZ masses are given, the second of which has the source removed. The X-ray masses are significantly above those found by weak lensing with a median ratio of 1.9.}
    \label{tab:full_cluster_list}\\
    %\begin{tabular}{lccccc} 
    Name & Map noise & X-ray mass & WL mass & tSZ mass & redshift \\
  & mJy/beam   & $\times10^{14}M_\odot$ & $\times10^{14}M_\odot$ & $\times10^{14}M_\odot$ &    \\\hline
  \endfirsthead
  \multicolumn{6}{c}{continued from previous page} \\\\
  Name & Map noise & X-ray mass & WL mass & tSZ mass & redshift \\
  & mJy/beam   & $\times10^{14}M_\odot$ & $\times10^{14}M_\odot$ & $\times10^{14}M_\odot$ &    \\\hline
  \endhead
eFEDS J082820.5$-$000722 & 0.10 & 6.78 & 3.52 & 2.41 & 0.84 \\
eFEDS J082955.4+004132 & 0.10 & 4.89 & 1.89 & 2.63 & 0.94 \\
eFEDS J083204.4+041908 & 0.09 & 2.56 & 1.57 & 1.37/2.13 & 0.20 \\
eFEDS J083250.0+031738 & 0.11 & 2.59 & 1.03 & 1.72 & 0.80 \\
eFEDS J083310.3+030137 & 0.12 & 2.25 & 0.94 & 0.21 & 0.75 \\
eFEDS J083509.0+001543 & 0.13 & 4.93 & 2.47 & 1.53 & 0.60 \\
eFEDS J083626.4+043822 & 0.12 & 2.71 & NA & 0.12 & 0.57 \\
eFEDS J083651.3+030002 & 0.13 & 2.70 & 1.46 & 1.38 & 0.20 \\
eFEDS J083713.3+034110 & 0.11 & 2.93 & NA & 0.10 & 1.00 \\
eFEDS J083834.2+020644 & 0.10 & 2.82 & 0.98 & 1.52 & 0.46 \\
eFEDS J083840.4+044417 & 0.12 & 2.00 & 0.83 & 0.10 & 0.45 \\
eFEDS J083857.6+020847 & 0.14 & 3.90 & 1.97 & 1.59/2.87 & 0.37 \\
eFEDS J083900.6+020057 & 0.10 & 2.80 & 1.60 & 0.86 & 0.35 \\
eFEDS J083929.7$-$015005$^*$ & 0.14 & 3.02 & 1.79 & 2.42/2.05 & 0.57 \\
eFEDS J083933.8$-$014044$^*$ & 0.15 & 5.15 & 4.90 & 2.59 & 0.28 \\
eFEDS J084006.2+025914 & 0.10 & 3.10 & 2.20 & 1.84 & 0.47 \\
eFEDS J084011.7+034837 & 0.11 & 3.11 & 1.42 & neg Y & 0.82 \\
eFEDS J084016.7+033952 & 0.09 & 2.85 & NA & neg Y & 0.81 \\
eFEDS J084021.7+020132 & 0.13 & 2.45 & NA & 1.04 & 0.38 \\
eFEDS J084035.8+044036 & 0.13 & 2.79 & NA & 0.26 & 0.79 \\
eFEDS J084044.7+024109 & 0.10 & 3.60 & 1.13 & 0.26 & 1.10 \\
eFEDS J084058.4+050408 & 0.13 & 4.26 & NA & neg Y & 1.26 \\
eFEDS J084105.6+031640 & 0.16 & 2.34 & 1.18 & 1.14 & 0.34 \\
eFEDS J084111.9$-$001947 & 0.13 & 4.98 & NA & 0.10 & 0.70 \\
eFEDS J084134.2+043504 & 0.18 & 2.52 & 1.15 & 0.73 & 0.56 \\
eFEDS J084146.5+045613 & 0.13 & 2.91 & 1.22 & 1.30 & 0.63 \\
eFEDS J084201.3+040534 & 0.18 & 2.48 & NA & neg Y & 0.76 \\
eFEDS J084223.1+003341$^*$ & 0.14 & 3.94 & 1.50 & 1.97 & 1.08 \\
eFEDS J084255.6+042333$^*$ & 0.14 & 3.03 & 2.45 & 2.12 & 0.48 \\
eFEDS J084434.3+031026 & 0.20 & 2.84 & 1.21 & 1.79 & 0.68 \\
eFEDS J084438.5+041946 & 0.19 & 2.43 & 1.07 & 1.11 & 0.58 \\
eFEDS J084439.0+043302 & 0.30 & 2.28 & NA & neg Y & 0.52 \\
eFEDS J084441.4+021702$^*$ & 0.15 & 4.00 & 2.15 & 2.53 & 0.66 \\
eFEDS J084501.0+012728$^*$ & 0.15 & 2.70 & 1.42 & 2.74 & 0.42 \\
eFEDS J084647.4+044608 & 0.19 & 2.28 & 1.82 & 0.10 & 0.24 \\
eFEDS J084717.8+033421 & 0.22 & 2.32 & 0.20 & 1.89 & 0.72 \\
eFEDS J084823.3+041205$^*$ & 0.14 & 4.44 & 1.87 & 2.61 & 0.87 \\
eFEDS J084833.2$-$012216$^*$ & 0.15 & 5.62 & 3.49 & 2.40 & 0.63 \\
eFEDS J084852.9+035940 & 0.20 & 2.74 & 1.64 & 0.10 & 0.67 \\
eFEDS J084910.6+024117 & 0.23 & 3.57 & 1.28 & 1.58 & 0.83 \\
eFEDS J084925.5+013841 & 0.10 & 4.02 & NA & neg Y & 0.51 \\
eFEDS J084939.6$-$005126$^*$ & 0.14 & 3.80 & 2.51 & 2.69 & 0.62 \\
eFEDS J084957.4+004524 & 0.13 & 4.18 & 2.29 & neg Y & 1.07 \\
eFEDS J085018.3+020019 & 0.18 & 2.07 & 0.92 & neg Y & 0.42 \\
eFEDS J085020.4+032820 & 0.13 & 2.09 & 0.96 & neg Y & 0.30 \\
eFEDS J085120.0+022952 & 0.16 & 2.97 & 1.56 & 1.78 & 0.38 \\
eFEDS J085217.0$-$010131$^*$ & 0.14 & 6.33 & 4.09 & 3.22 & 0.47 \\
eFEDS J085447.1$-$012133$^*$ & 0.14 & 3.31 & 2.16 & 2.48 & 0.35 \\
eFEDS J085530.1$-$010635$^*$ & 0.15 & 5.66 & 3.89 & 2.69 & 0.75 \\
eFEDS J085542.7+032807 & 0.17 & 2.50 & 1.26 & 0.86 & 0.74 \\
eFEDS J085547.0+025458 & 0.21 & 3.78 & 1.73 & neg Y & 0.97 \\
eFEDS J085620.8+014650$^*$ & 0.15 & 5.90 & 3.04 & 3.64 & 0.73 \\
eFEDS J085627.2+014218$^*$ & 0.14 & 8.59 & 5.74 & 2.84 & 0.73 \\
eFEDS J085635.0+031342 & 0.14 & 2.10 & 0.92 & 0.54 & 0.39 \\
eFEDS J085751.7+031039$^*$ & 0.14 & 5.63 & 4.48 & 4.58 & 0.20 \\
eFEDS J085901.1$-$012026$^*$ & 0.14 & 4.66 & 1.39 & 1.86 & 1.30 \\
eFEDS J085931.9+030839$^*$ & 0.15 & 2.81 & 1.58 & 1.81 & 0.20 \\
eFEDS J090004.4+033325 & 0.14 & 2.78 & NA & 0.10 & 0.70 \\
eFEDS J090033.8+033933 & 0.08 & 4.46 & 1.87 & 1.41 & 0.81 \\
eFEDS J090059.3+035926 & 0.14 & 3.12 & 1.53 & 1.47 & 0.42 \\
eFEDS J090131.2+030057$^*$ & 0.15 & 3.50 & 1.59 & 2.08/2.13 & 0.19 \\
eFEDS J090144.8+040827$^*$ & 0.16 & 4.25 & 2.56 & 2.13 & 0.84 \\
eFEDS J090255.6+044036 & 0.16 & 2.07 & 1.00 & 0.87 & 0.45 \\
eFEDS J090323.7+030738 & 0.15 & 2.96 & 1.86 & 1.24 & 0.41 \\
eFEDS J090328.7$-$013622$^*$ & 0.17 & 4.40 & 4.29 & 2.83 & 0.44 \\
eFEDS J090336.8+033125 & 0.14 & 2.94 & 1.36 & neg Y & 0.62 \\
eFEDS J090417.1+040439 & 0.15 & 2.40 & NA & 0.78 & 0.54 \\
eFEDS J090418.7+020642$^*$ & 0.17 & 3.93 & 1.62 & 2.35 & 0.81 \\
eFEDS J090430.8+042649$^*$ & 0.19 & 3.16 & 1.40 & 2.43 & 0.47 \\
eFEDS J090452.5+033326 & 0.11 & 4.29 & 2.26 & 2.04 & 0.81 \\
eFEDS J090540.1+043441$^*$ & 0.23 & 4.02 & 2.88 & 1.81/2.63 & 0.24 \\
eFEDS J090540.8+013220 & 0.12 & 4.24 & 1.71 & 1.68 & 0.66 \\
eFEDS J090614.7$-$010820 & 0.11 & 4.63 & NA & neg Y & 0.89 \\
eFEDS J090627.5+035846 & 0.15 & 2.37 & 1.37 & 1.54 & 0.48 \\
eFEDS J090634.9+045034 & 0.11 & 4.52 & NA & neg Y & 0.88 \\
eFEDS J090637.0+010852 & 0.10 & 4.35 & 2.09 & 1.28 & 0.79 \\
eFEDS J090718.7+035258 & 0.12 & 3.50 & 1.91 & 1.98 & 0.73 \\
eFEDS J090754.5+005738$^*$ & 0.20 & 3.39 & 2.09 & 2.45 & 0.74 \\
eFEDS J090757.6+025428$^*$ & 0.21 & 2.82 & 1.19 & 2.06 & 0.81 \\
eFEDS J090803.9+020046 & 0.14 & 2.64 & 1.41 & 1.07 & 0.47 \\
eFEDS J090806.5+032613 & 0.11 & 2.53 & NA & 0.10 & 0.74 \\
eFEDS J090821.9+025141 & 0.12 & 3.87 & 1.43 & 1.72 & 0.81 \\
eFEDS J090849.7+042242 & 0.12 & 5.43 & 2.41 & 2.03 & 0.80 \\
eFEDS J090915.3$-$010104 & 0.11 & 4.65 & 3.04 & 1.47 & 0.82 \\
eFEDS J090930.6+034056 & 0.11 & 4.44 & 2.96 & neg Y/2.47 & 0.74 \\
eFEDS J090932.6$-$005020$^*$ & 0.21 & 4.74 & 1.53 & 2.09 & 1.13 \\
eFEDS J091057.2+041730 & 0.13 & 2.39 & 1.26 & 1.20 & 0.47 \\
eFEDS J091111.1+040016 & 0.16 & 2.36 & 1.24 & 0.10 & 0.50 \\
eFEDS J091135.9+034627 & 0.10 & 4.09 & NA & neg Y & 0.93 \\
eFEDS J091254.5+032029 & 0.12 & 4.89 & 2.59 & 2.02 & 0.62 \\
eFEDS J091315.0+034850$^*$ & 0.16 & 4.16 & 2.48 & 2.99 & 0.44 \\
eFEDS J091358.2+025707 & 0.11 & 2.51 & 1.24 & neg Y & 0.43 \\
eFEDS J091433.9+022718 & 0.11 & 2.01 & 1.03 & neg Y/neg Y & 0.32 \\
eFEDS J091446.2+001048$^*$ & 0.12 & 3.22 & 3.18 & 2.15 & 0.52 \\
eFEDS J091509.5+051521 & 0.11 & 2.78 & 1.97 & 0.10 & 0.25 \\
eFEDS J091522.5+041201 & 0.13 & 2.11 & 1.22 & 1.03 & 0.46 \\
eFEDS J091555.7$-$013248$^*$ & 0.13 & 6.80 & 4.69 & 2.76 & 0.50 \\
eFEDS J091610.1$-$002349$^*$ & 0.13 & 6.24 & 3.69 & 4.33 & 0.32 \\
eFEDS J091647.0+015532$^*$ & 0.14 & 1.81 & 0.95 & 2.35 & 0.27 \\
eFEDS J091648.2+030506 & 0.17 & 2.77 & 1.66 & neg Y & 0.62 \\
eFEDS J091849.0+021205$^*$ & 0.12 & 2.92 & 1.26 & 2.11/2.20 & 0.28 \\
eFEDS J091850.7+030943 & 0.11 & 3.21 & NA & 1.17 & 0.84 \\
eFEDS J091900.1+035311 & 0.51 & 2.27 & 1.05 & 0.00 & 0.40 \\
eFEDS J091957.8+035013 & 0.12 & 3.66 & 2.11 & 1.09 & 0.52 \\
eFEDS J092023.2+013444$^*$ & 0.12 & 6.37 & 3.67 & 2.79 & 0.71 \\
eFEDS J092031.9+040621 & 0.12 & 2.58 & 1.13 & 1.24 & 0.60 \\
eFEDS J092041.2+041118 & 0.15 & 2.12 & 0.97 & 0.10 & 0.58 \\
eFEDS J092049.5+024514 & 0.12 & 3.54 & 2.99 & 1.30 & 0.28 \\
eFEDS J092121.2+031726$^*$ & 0.13 & 6.75 & 5.40 & 4.19 & 0.35 \\
eFEDS J092209.4+034629$^*$ & 0.13 & 4.53 & 2.51 & 1.86 & 0.27 \\
eFEDS J092212.1$-$002731$^*$ & 0.13 & 4.37 & 2.31 & 2.99 & 0.32 \\
eFEDS J092258.2+032042 & 0.12 & 2.51 & 1.49 & 0.94 & 0.57 \\
eFEDS J092328.3+043107 & 0.13 & 2.85 & 1.05 & 0.53 & 0.66 \\
eFEDS J092402.3+054206 & 0.10 & 5.81 & 4.03 & 1.41 & 0.49 \\
eFEDS J092619.9+021208 & 0.44 & 2.91 & NA & 0.00/3.04 & 0.49 \\
eFEDS J092631.1+014512 & 0.12 & 4.88 & 2.82 & 1.58 & 0.76 \\
eFEDS J092650.5+035755 & 0.14 & 2.70 & 1.82 & 0.10 & 0.37 \\
eFEDS J092737.8+020608 & 0.14 & 4.29 & 2.19 & 1.72 & 0.42 \\
eFEDS J092821.2+042149 & 0.12 & 2.61 & 1.39 & 0.90 & 0.23 \\
eFEDS J092832.5+041517 & 0.14 & 2.14 & 1.09 & neg Y & 0.44 \\
eFEDS J092859.1+040532 & 0.47 & 2.29 & 1.10 & 0.36 & 0.26 \\
eFEDS J092910.2+022034 & 0.64 & 2.56 & 1.45 & 0.92 & 0.44 \\
eFEDS J092918.4+044925 & 0.14 & 2.08 & 1.05 & 0.10 & 0.35 \\
eFEDS J092921.8+040040 & 0.13 & 4.21 & 2.18 & 1.45/2.96 & 0.50 \\
eFEDS J093011.3+031648 & 0.12 & 3.71 & 1.17 & 0.64 & 0.76 \\
eFEDS J093025.3+021714$^*$ & 0.11 & 5.32 & 3.17 & 4.11 & 0.55 \\
eFEDS J093301.4+024301 & 0.13 & 2.64 & NA & 0.10 & 0.84 \\
eFEDS J093332.2+020934 & 0.15 & 2.78 & NA & 1.25 & 0.84 \\
eFEDS J093500.8+005417$^*$ & 0.12 & 3.69 & 2.33 & 2.79 & 0.38 \\
eFEDS J093513.1+004758$^*$ & 0.12 & 7.62 & 6.68 & 5.31 & 0.36 \\
eFEDS J093521.0+023234$^*$ & 0.12 & 7.55 & 2.55 & 4.40 & 0.51 \\
eFEDS J093522.2+032329 & 0.15 & 3.31 & 1.76 & 0.74 & 0.34 \\
eFEDS J093525.9+035101 & 0.15 & 3.72 & NA & neg Y & 1.26 \\
eFEDS J093630.9+031838 & 0.11 & 2.20 & 0.91 & 0.10 & 0.41 \\
eFEDS J093742.8+033842 & 0.17 & 2.83 & 1.44 & 1.08 & 0.61 \\
eFEDS J093938.3+042218 & 0.20 & 2.95 & 1.62 & 1.50 & 0.37 \\
eFEDS J094005.9+031329 & 0.12 & 2.98 & 1.61 & 0.98 & 0.48 \\
eFEDS J094007.4+035755 & 0.13 & 2.78 & 1.21 & 1.52 & 0.67 \\

%\end{tabular}
    
\end{longtable}

\end{document}